\documentclass[twocolumn,trackchanges]{aastex62}

\usepackage{subfigure}
\usepackage{color}
\usepackage{comment}
\usepackage{url}

\newcommand{\etaTel}{$\eta$~Tel}

\newcommand{\OI}{\ion{O}{1}}
\newcommand{\CI}{\ion{C}{1}}
\newcommand{\CII}{\ion{C}{2}}
\newcommand{\NI}{\ion{N}{1}}
\newcommand{\SII}{\ion{S}{2}}
\newcommand{\SI}{\ion{S}{1}}
\newcommand{\FeII}{\ion{Fe}{2}}
\newcommand{\AlII}{\ion{Al}{2}}
\newcommand{\SiII}{\ion{Si}{2}}
\newcommand{\MnII}{\ion{Mn}{2}}
\newcommand{\MgII}{\ion{Mg}{2}}
\newcommand{\kms}{km s$^{-1}$}

\begin{document}

\title{A Radiatively Driven Wind from the \etaTel\ Debris Disk}

\author[0000-0002-1176-3391]{Allison Youngblood}
\affiliation{Exoplanets and Stellar Astrophysics Lab, NASA Goddard Space Flight Center, Greenbelt, MD 20771, USA}
\affiliation{Laboratory for Atmospheric and Space Physics, University of Colorado, 600 UCB, Boulder, CO 80309, USA}
\email{allison.a.youngblood@nasa.gov}

\author[0000-0002-2989-3725]{Aki Roberge}
\affiliation{Exoplanets and Stellar Astrophysics Lab, NASA Goddard Space Flight Center, Greenbelt, MD 20771, USA}

\author[0000-0001-7891-8143]{Meredith A. MacGregor} 
\affiliation{Center for Astrophysics and Space Astronomy, University of Colorado, 389 UCB Boulder, CO 80309, USA}

\author[0000-0002-7201-7536]{Alexis Brandeker}
\affiliation{AlbaNova University Centre, Stockholm University, Department of Astronomy, Stockholm, Sweden}

\author{Alycia Weinberger} 
\affiliation{Department of Terrestrial Magnetism, Carnegie Institution for Science, 5241 Broad Branch Road NW, Washington, DC 20015, USA}

\author[0000-0003-2953-755X]{Sebasti\'an P\'erez}
\affiliation{Departamento de F\'isica, Universidad de Santiago de Chile. Avenida Ecuador 3493, Estaci\'on Central, Santiago, Chile}
\affiliation{Center for Interdisciplinary Research in Astrophysics and Space Exploration (CIRAS), Universidad de Santiago de Chile, Chile}

\author{Carol Grady}
\affiliation{Eureka Scientific, 2452 Delmer, Suite 100, Oakland, CA 96002, USA}

\author{Barry Welsh}
\affiliation{Eureka Scientific, 2452 Delmer, Suite 100, Oakland, CA 96002, USA}

\begin{abstract}

We present far- and near-ultraviolet absorption spectroscopy of the $\sim$23 Myr edge-on debris disk surrounding the A0V star $\eta$~Telescopii, obtained with the \emph{Hubble Space Telescope} Space Telescope Imaging Spectrograph. We detect absorption lines from \CI, \CII, \OI, \MgII, \AlII, \SiII, \SII, \MnII, \FeII, and marginally \ion{N}{1}. The lines show two clear absorption components at $-22.7\pm0.5$ \kms\ and $-17.8\pm0.7$ \kms, which we attribute to circumstellar (CS) and interstellar (IS) gas, respectively. CO absorption is not detected, and we find no evidence for star-grazing exocomets. The CS absorption components are blueshifted by $-16.9\pm$2.6 \kms\ in the star's reference frame, indicating that they are outflowing in a radiatively driven disk wind. We find that the C/Fe ratio in the \etaTel\ CS gas is significantly higher than the solar ratio, as is the case in the $\beta$~Pic and 49 Cet debris disks. Unlike those disks, however, the measured C/O ratio in the \etaTel\ CS gas is consistent with the solar value. Our analysis shows that because \etaTel\ is an earlier type star than $\beta$~Pic and 49 Cet, with more substantial radiation pressure at the dominant \CII\ transitions, this species cannot bind the CS gas disk to the star as it does for $\beta$~Pic and 49 Cet, resulting in the disk wind.

\end{abstract}


\section{Introduction} \label{sec:Introduction}

Debris disks are the end stage of planet formation, maintained by colliding planetesimals that persist beyond the gas-rich protoplanetary disk phase. The dust properties of these systems have been extensively observed, but their gas properties are woefully under-constrained \citep[][and references therein]{Hughes:2018}. Sub-mm CO emission has been detected in only about 20 debris disks to date \citep[e.g.,][]{Zuckerman:1995,Dent:2014,Moor:2017}. Any gas within a debris disk is unlikely to be a remnant from the protoplanetary stage unless the system is especially young \citep[e.g.,][]{Kral:2019}. Instead, the preferred explanation is that debris disk gas is continually replenished through various processes ultimately associated with destruction of planetesimals \citep{Beust:1990,Czechowski:2007,Grigorieva:2007,Zuckerman:2012}. Thus, examining the gas in debris disks offers the chance to probe the chemical composition of the young extrasolar planetesimals themselves.

For edge-on debris disks, small amounts of circumstellar (CS) gas are detectable with line-of-sight absorption spectroscopy, where the central star serves as the background source. This method has been successfully applied to numerous debris disks in the optical \citep[e.g.,][]{Hobbs:1985,Montgomery:2012,Rebollido:2018,Rebollido:2020} and ultraviolet \citep[e.g.,][]{Roberge2006,Roberge2014,Jenkins2020_51Oph}. The gas seen has low column densities ($N \gtrsim 10^{11}$ cm$^{-2}$) and appears to be primarily in atomic and ionic forms. The relative paucity of molecular gas is a consequence of the low dust densities \citep[e.g.][]{Chen:2014} and the apparent lack of H$_2$ in debris disks \citep[e.g.,][]{LecavelierdesEtangs:2001,Chen:2007}, which make them generally optically thin to dissociating interstellar radiation \citep[e.g.,][]{Matra:2017}. However, in the few debris disks displaying substantial CO emission, ionization of neutral carbon may play a role in shielding the molecule from rapid dissociation \citep{Kral:2019}.

In the case of $\beta$~Pic, the presence of a stable (i.e., unshifted with respect to the star) gas disk in Keplerian rotation was a puzzle, as many of the species observed should be rapidly blown out by radiation pressure from the central star \citep{Lagrange:1998}. This mystery was resolved by incorporating Coulomb effects in dynamical models of the gas disk \citep{Fernandez:2006} and observing that the $\beta$~Pic gas was extremely overabundant in carbon relative to solar abundance \citep{Roberge2006,Cataldi:2014}. The ionized gas is dynamically coupled into a single fluid with an abundance-weighted, effective radiation pressure coefficient. Since ionized carbon feels weak radiation pressure from an A5V star, the carbon overabundance lowers the fluid's radiation pressure coefficient enough such that the whole fluid is bound to the star. Neutral species that feel strong radiation pressure are rapidly ionized and join the fluid. This is likely also the explanation for the presence of stable atomic/ionic gas in the 49~Cet debris disk \citep{Roberge2014}. However, a comprehensive understanding of the coupled dynamics and composition of gas in debris disks remains elusive.

\object[HD 181296]{Eta Telescopii} (HD~181296, HR 7329) is a young A0V star at 47.36~pc \citep{GaiaDR2} with a M7/8V companion $\sim$4\arcsec\ away \citep{Lowrance2000,Guenther2001}. A debris disk identified via infrared excess \citep{Backman1993,Mannings1998} surrounds the A0V star (\etaTel\ A). Hereafter we refer to \etaTel~A as simply \etaTel. The star is a member of the $\beta$~Pic Moving Group and therefore its age is $23 \pm 3$~Myr \citep{Zuckerman2001,Mamajek2014}. 
Other fundamental stellar properties of \etaTel\ include $T_\mathrm{eff} = 9500-10400$~K \citep{McCarthy2012,Rebollido:2018}, $R = 1.61~ R_{\odot}$ \citep{Rhee2007}, and $v \sin i = 230$ \kms\ \citep{Rebollido:2018}. The metallicity [Fe/H] appears to be consistent with solar \citep{Saffe2008}.

\textit{Spitzer} IRS spectroscopy of \etaTel\ revealed emission consistent with CS dust at two temperatures (115 and 370 K), indicating the presence of two different planetesimal belts, akin to the asteroid and Kuiper belts of the solar system \citep{Chen2006}. \cite{Smith2009} resolved the debris disk with Gemini South T-ReCS imaging at 18.3~$\mu$m, discovering its edge-on orientation and confirming the presence of two distinct temperature components: a spatially resolved cold component at 24~au and an unresolved hot component at $\sim$4~au. Given the age of the system, the source of the infrared-emitting dust grains is likely collisions between planetesimals.

The only CS gas tracer definitively detected to date in \etaTel's debris disk is far-IR \ion{C}{2} emission seen with the \textit{Herschel} PACS instrument \citep{RiviereMarichalar2014}. With a system age of $\sim$23 Myr, the \ion{C}{2} gas is not expected to be a remnant of the protoplanetary disk. No \ion{O}{1} or H$_2$O emission from \etaTel\ was detected with \textit{Herschel} \citep{Dent2013}, and CO emission was not detected with APEX \citep{Hales2014}. Optical \ion{Ca}{2} and \ion{Na}{1} absorption lines have been detected towards \etaTel\ \citep{Welsh2018,Rebollido:2018,Rebollido:2020}, but their origin could not be definitively identified as CS or interstellar (IS).
\cite{Welsh2018} searched for signs of exocomets in optical spectra of \etaTel; some weak time-variable \ion{Ca}{2} absorption was seen but could not be confidently identified as arising from star-grazing exocomets.

In this paper, we present a sensitive search for CS absorption from \etaTel's edge-on debris disk using two epochs of ultraviolet (UV) data obtained with the \emph{Hubble Space Telescope} (\emph{HST}) Space Telescope Imaging Spectrograph (STIS) instrument. Section~\ref{sec:ObservationsReductions} describes the observations and data reduction. In Section~\ref{sec:methods}, we detail how we determined the physical properties of the detected UV absorption lines, including radial velocities and column densities, and identify the origins of the absorption components seen. Section~\ref{sec:methods} also presents new upper limits on CO using data from both \emph{HST} and the Atacama Large Millimeter/submillimeter Array (ALMA), and discusses an unidentified time-variable absorption feature seen in the \emph{HST} spectra. Section~\ref{section:velocity} discusses the velocity structure of the line of sight to \etaTel, including a determination of the stellar radial velocity. Our analyses of the elemental abundances in and the dynamics of the \etaTel\ CS gas appear in Sections~\ref{section:abundances} and \ref{section:dynamics}, respectively.
Section~\ref{section:discussion} discusses our interpretation of the CS gas and concluding remarks appear in Section~\ref{sec:conclusion}.

\section{Observations and Data Reduction} \label{sec:ObservationsReductions}

\begin{figure*}
\includegraphics[width=\textwidth]{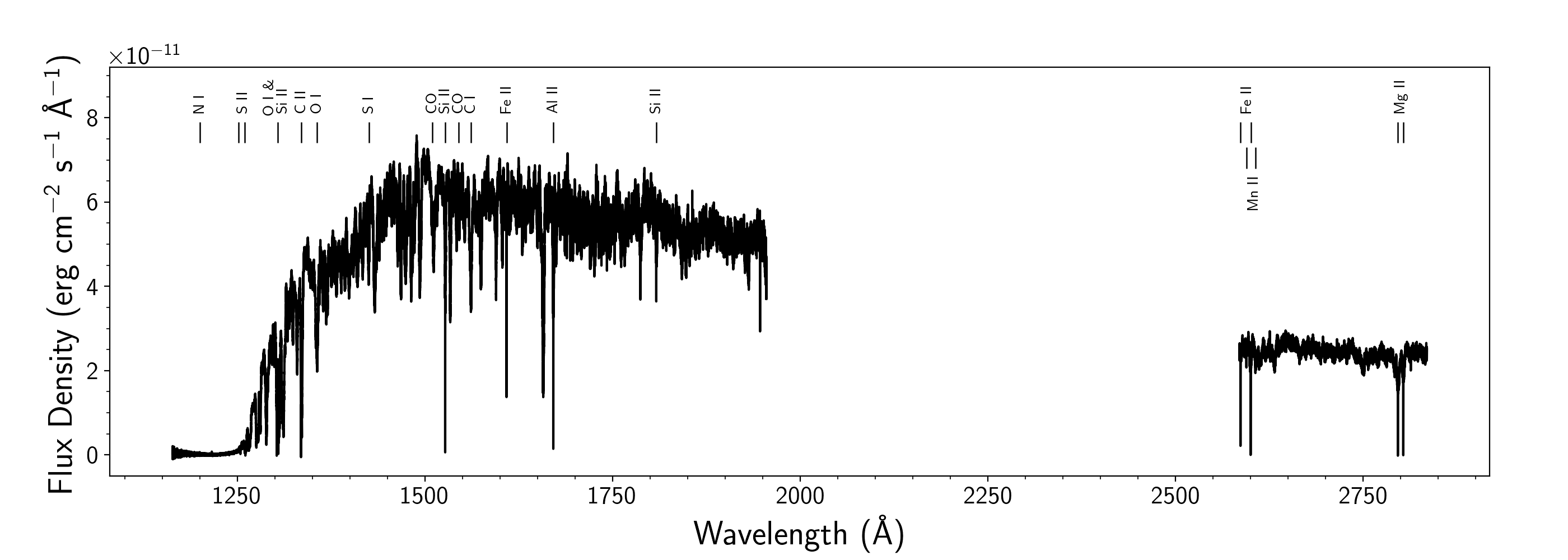}
\caption{The co-added spectrum of \etaTel~from HST STIS E140H and E230H observations. The atomic and molecular transitions analyzed in this work are labeled.} \label{fig:all_spectra}
\end{figure*}

We observed the \etaTel~debris disk with the \emph{HST} STIS spectrograph on 2016-Oct-03 (Visit~1) and 2016-Oct-06 (Visit~2). The spectra were acquired in two visits to look for changes in CS absorption lines caused by star-grazing comets, which are observed in spectra of numerous young debris disks \citep[see][]{Rebollido:2020}. The wavelength coverage of the STIS spectra is 
1164--1951~\AA\ and 2587--2851~\AA; the complete spectrum combining both visits appears in Figure~\ref{fig:all_spectra}. The spectra were obtained with the E140H and E230H echelle gratings (grating settings i1271, c1416, c1598, i1813, and i2713), providing an average dispersion of $\Delta \lambda = \lambda / 228\,000$~\AA~pixel$^{-1}$. The data were calibrated with the default CALSTIS v3.4.2 pipeline. The relative wavelength accuracy within an exposure is better than $0.5~\mathrm{pixel} = 0.66 \ \mathrm{km} \ \mathrm{s}^{-1}$ and the absolute wavelength accuracy across exposures is better than $1~\mathrm{pixel} = 1.32 \ \mathrm{km} \ \mathrm{s}^{-1}$.\footnote{STIS Instrument Handbook, version 19.0: \url{https://hst-docs.stsci.edu/stisihb}.} 

No significant variability in the spectra was seen, with the exception of one mysterious absorption feature (discussed in Section~\ref{subsec:Mystery}) and a small $\sim$8\% flux offset between the c1598 spectra (1497-1699 \AA) from Visit~1 and Visit~2. As the spectra from the other grating settings did not show this flux variation between visits, this inter-visit variation in the c1598 spectra is very likely a calibration error, possibly due to time-dependent shifts in the echelle blaze function. We did not correct this flux offset, as our absorption line analysis is performed relative to the continuum flux.

We created a master \etaTel\ spectrum by combining the individual spectra from both visits, first interpolating all onto a common wavelength grid and then performing a weighted average. The first and last 10 grid points of each E140H echelle order were excised to avoid introduction of spurious features by edge effects; the edge effects in the E230H spectra were not as consistent from order to order so we manually determined the number of edge grid points to excise from each order. The error arrays were added in quadrature and divided by the number of spectra averaged at that wavelength. We manually estimated and subtracted additional inter-order scattered light for strongly saturated absorption features whose troughs should reach zero flux (\OI~1302 \AA~and \CII~1334 \AA). The scattered light corrections are small and do not have a significant impact on the quality of our absorption feature modeling.

The spectra show many narrow absorption lines arising from gas along the line of sight to the central star. The species detected are: \CI, \CII, \NI, \OI, \MgII, \AlII, \SiII, \SII, \MnII, and \FeII. For all lines analyzed, the rest wavelengths ($\lambda_0$) and the lower energy levels of the transitions ($E_l$) appear in Table~\ref{table:lines}. With the exception of the aforementioned ``mystery feature'' (Section~\ref{subsec:Mystery}), none of the absorption lines varied between the visits; thus, there is no clear evidence of star-grazing planetesimals in the data. We find no emission lines in the STIS spectra, indicating that \etaTel\ is chromospherically inactive, as expected for stars with spectral type earlier than $\sim$A5V \citep[][and references therein]{Braithwaite2017}.

\startlongtable
\begin{deluxetable*}{lCRcCC}
\tablecolumns{6}
\tablewidth{0pt}
\tablecaption{Absorption line analysis \label{table:lines}} 
\tablehead{\colhead{$\lambda_0$} &
           \colhead{$E_\mathrm{l}$} &
           \colhead{$v_\mathrm{r}$} &
           \colhead{Component} &
           \colhead{$\log_{10} N_l$} &
           \colhead{$b$} \\
           \colhead{(\AA)} & 
           \colhead{cm$^{-1}$} &
           \colhead{(\kms)} & 
           \colhead{ID} & 
           \colhead{(cm$^{-2}$)} &
           \colhead{(\kms)} 
}
\startdata
\sidehead{\textbf{\CI} $(\beta = 51.7)$}
1560.3092    & 0.0 & $-24.06^{+2.12}_{-1.41}$ & CS & $11.71^{+0.23}_{-0.76}$ & $3.01^{+0.68}_{-1.08}$ \\
1560.6822, 1560.7090 & 16.40 & $-24.06^{+2.12}_{-1.41}$ & CS & $11.15^{+0.63}_{-2.09}$ & $3.01^{+0.68}_{-1.08}$ \\
1561.3402, 1561.3667, 1561.4384 & 43.40 & $-24.06^{+2.12}_{-1.41}$ & CS & $11.46^{+0.29}_{-2.09}$ & $3.01^{+0.68}_{-1.08}$ \\
\multicolumn{3}{l}{Total in all energy levels} & CS & $11.97^{+0.33}_{-1.01}$ & \\
\hline
\sidehead{\textbf{\CII} $(\beta = 19.3)$}
1334.5323     & 0.0 & $-22.53^{+1.33}_{-1.33}$ & CS & $16.16^{+0.18}_{-0.36}$ & $3.11^{+0.17}_{-0.16}$ \\
              &  & $-18.60^{+1.35}_{-1.30}$ & IS & $15.50^{+0.14}_{-0.13}$ & 
$5.88^{+0.19}_{-0.20}$ \\
\textbf{1335.6627,} 1335.7077     & 63.42 & $-22.53^{+1.33}_{-1.33}$ & CS & $13.05^{+0.02}_{-0.03}$ & $3.11^{+0.17}_{-0.16}$ \\
\multicolumn{3}{l}{Total in all energy levels} & CS & $16.17^{+0.18}_{-0.35}$ & \\ 
\hline
\sidehead{\textbf{\NI} $(\beta = 0.09)$}
1199.5496, 1200.2233, 1200.7098 & 0.0 & -21.06$^{+1.57}_{-1.64}$ & CS & $13.53-17.24$ \ \tablenotemark{a} & $2.76^{+1.14}_{-0.69}$ \\
\multicolumn{3}{l}{Total in all energy levels} & CS & $13.53-17.24$ \  & \\
\hline
\sidehead{\textbf{\OI} $(\beta = 1.0)$}
1302.1685, 1355.5977   & 0.0 & $-25.27^{+1.33}_{-1.32}$ & CS & $16.17^{+0.21}_{-0.30}$ & $1.82^{+0.10}_{-0.12}$ \\
             &     & $-17.56^{+1.54}_{-1.45}$ & IS & $14.68^{+0.09}_{-0.07}$  & $7.97^{+0.41}_{-0.64}$ \\
             &     & $-0.76^{+2.25}_{-4.91}$ & unknown & $13.47^{+0.24}_{-0.19}$ & $8.17^{+3.51}_{-2.14}$ \\
1304.8576    & 158.27 & $-25.27^{+1.33}_{-1.32}$ & CS & $13.32^{+0.03}_{-0.03}$ &  $1.82^{+0.10}_{-0.12}$ \\
1306.0286    & 226.98 & $-25.27^{+1.33}_{-1.32}$ & CS & $12.85^{+0.05}_{-0.05}$  &  $1.82^{+0.10}_{-0.12}$ \\
\multicolumn{3}{l}{Total in all energy levels} & CS & $16.17^{+0.21}_{-0.30}$ & \\
\hline
\sidehead{\textbf{\MgII} $(\beta = 365.7)$}
2796.352, 2803.531 & 0.0 & -21.19$^{+1.32}_{-1.32}$ & CS & 14.16$^{+0.08}_{-0.07}$ & 4.00$^{+0.10}_{-0.11}$ \\
 &  & -10.16$^{+1.31}_{-1.34}$ & IS & 12.26$^{+0.03}_{-0.02}$ & 2.71$^{+0.17}_{-0.15}$ \\
\multicolumn{3}{l}{Total in all energy levels} & CS & 14.16$^{+0.08}_{-0.07}$ &  \\
\hline
\sidehead{\textbf{\AlII} $(\beta = 190.7)$}
1670.7874 & 0.0 & $-22.56^{+1.36}_{-1.28}$ & CS & $12.44^{+0.04}_{-0.02}$ & $2.58^{+0.23}_{-0.13}$ \\
          &     & $-18.67^{+2.66}_{-1.28}$ & IS & $11.71^{+0.08}_{-0.23}$ & $6.20^{+0.85}_{-1.36}$ \\
\multicolumn{3}{l}{Total in all energy levels} & CS & $12.44^{+0.04}_{-0.02}$ & \\
\hline
\sidehead{\textbf{\SiII} $(\beta = 32.2)$}
1304.3742, 1526.7066, 1808.0129 & 0.0 & $-22.26^{+1.31}_{-1.33}$ & CS & $14.04^{+0.03}_{-0.03}$  & $3.05^{+0.08}_{-0.09}$ \\
            & & $-17.37^{+1.40}_{-1.38}$ & IS & $13.18^{+0.05}_{-0.05}$  & $6.05^{+0.24}_{-0.27}$ \\
\multicolumn{3}{l}{Total in all energy levels \tablenotemark{b}} & CS & $14.04^{+0.03}_{-0.03}$ & \\
\hline
\sidehead{\textbf{\ion{S}{1}}$(\beta = 35.0)$}
1425.0299, 1425.1877, 1425.2189 & 0.0 &  & CS & $< 12.07$ \ \tablenotemark{c} & \\
\multicolumn{3}{l}{Total in all energy levels \tablenotemark{b}} & CS & $< 12.07^{ \ \, }$ & \\
\hline
\sidehead{\textbf{\SII} $(\beta = 0.05)$} 
1250.578, 1253.805, 1259.518 & 0.0 & $-21.32^{+1.42}_{-1.37}$ & CS & $14.35^{+0.03}_{-0.06}$ & $3.19^{+0.24}_{-0.25}$ \\
            &   & $-17.37^{+4.99}_{-2.06}$& IS & $12.58^{+1.33}_{-3.15}$ & $4.75^{+6.75}_{-4.60}$ \\
\multicolumn{3}{l}{Total in all energy levels} & CS & $14.35^{+0.03}_{-0.06}$ & \\
\hline
\sidehead{\textbf{\MnII} $(\beta = 82.7)$} 
2594.499, 2606.462 & 0.0 & -22.69$^{+1.33}_{-1.33}$ & CS & 11.76$^{+0.03}_{-0.03}$ & 2.90$^{+0.27}_{-0.25}$ \\
 &  &  & IS & $<$11.28 \ \tablenotemark{d} & \\
\multicolumn{3}{l}{Total in all energy levels} & CS & 11.76$^{+0.03}_{-0.03}$ & \\
\hline
\sidehead{\textbf{\FeII} $(\beta = 79.3)$} 
1608.4511, 2586.6500, 2600.1729 & 0.0 & $-22.79^{+1.32}_{-1.32}$ & CS & $13.48^{+0.01}_{-0.01}$ & $2.59^{+0.04}_{-0.04}$ \\
          &     & $-17.64^{+1.44}_{-1.38}$ & IS & $12.64^{+0.04}_{-0.04}$ & $6.99^{+0.35}_{-0.39}$ \\
\multicolumn{3}{l}{Total in all energy levels \tablenotemark{b}} & CS & $13.48^{+0.01}_{-0.01}$ & \\
\hline
\sidehead{\textbf{CO} \tablenotemark{e} $(\beta \leq 11.6)$} 
$J_l = 0$ & 0.0 &  & CS & $<12.38$ \ \tablenotemark{f} & \\
$J_l = 1$ & 3.85 &  & CS & $<12.47$ \ \tablenotemark{f} & \\
$J_l = 2$ & 11.53 &  & CS & $<12.66$ \ \tablenotemark{f} & \\
$J_l = 3$ & 23.07 &  & CS & $<12.95$ \ \tablenotemark{f} & \\
$J_l = 4$ & 38.45 &  & CS & $<12.15$ \ \tablenotemark{f} & \\
$J_l = 5$ & 57.67 &  & CS & $<12.09$ \ \tablenotemark{f} & \\
$J_l = 6$ & 80.74 &  & CS & $<12.23$ \ \tablenotemark{f} & \\
Total &  &  & CS & $<13.05$ \ \tablenotemark{g} &  \\ 
\enddata
\tablecomments{Uncertainties represent the 68\% confidence interval. All radial velocities ($v_{\rm r}$) are in the heliocentric frame and have had the STIS absolute wavelength accuracy value (1.32 \kms) propagated into their confidence intervals.  In the rows labeled ``Total in all energy levels'', the column density listed is the total of the column densities in all the fine structure energy levels of the ground term.}
\tablenotetext{a}{Lower limit and upper limit (both $3\sigma$) reported. The marginalized posterior distribution for the column density parameter is wide and flat-topped with a median value of 15.41 and a $1\sigma$~range of $14.42-16.41$.} 
\tablenotetext{b}{Lines arising from excited fine structure energy levels of the ground term not seen in the spectra. Total column density in all levels is assumed to be equal to the column density in the ground energy level.}
\tablenotetext{c}{Upper limit ($3\sigma$) computed with $v_{\rm r}$ allowed to vary between $-26$ and $-21$~\kms\ and $b$ allowed to vary between 0.1 and 10 \kms.} 
\tablenotetext{d}{Upper limit ($3\sigma$) computed with $v_{\rm r}$ allowed to vary between $-20$ and $-10$~\kms\ and $b$ allowed to vary between 0.1 and 15 \kms.} 
\tablenotetext{e}{See Table 3A of \cite{Morton1994} for a list of the CO transition wavelengths associated with each $J_l$ rotational level of the lowest vibrational state ($\nu$=0) of the ground electronic state ($X$).}
\tablenotetext{f}{Upper limit ($3\sigma$) on CO column density in each $J_l$ rotational level of the lowest 
vibrational state ($\nu = 0$) of the ground electronic state ($X$).}
\tablenotetext{g}{Upper limit ($3\sigma$) on total CO column density from fitting the $A-X \ (0-0)$ and $A-X \ (1-0)$ bands simultaneously assuming a Boltzmann distribution for the rotational energy levels. Excitation temperature was allowed to vary between 1 and 500~K.}
\end{deluxetable*}

\section{Fitting the absorption features} \label{sec:methods}

\subsection{Atomic and Ionic Species}

To measure column densities and radial velocities of the CS absorbers, we created models of the observed absorption features in the following way. For each species listed in Table~\ref{table:lines} (\CI, \CII, \OI, etc.), absorption lines arising from the different lower energy levels ($E_l$) were each modeled as a Voigt line profile. The parameters for each Voigt profile are radial velocity ($v_{\rm r}$), the line of sight column density in the lower energy level of the transition ($N_l$), and the Doppler broadening parameter ($b$). The vacuum rest wavelengths, oscillator strengths, and radiation damping constants used in constructing the Voigt profiles are from \cite{Morton1994} and \cite{Morton2003}. For each transition that might contain an IS absorption component, an additional Voigt profile with an independent radial velocity was included. 

The resulting ensemble of Voigt profiles was convolved with the instrumental line spread function for the appropriate grating and slit\footnote{STIS Instrument Handbook, version 19.0: \url{https://hst-docs.stsci.edu/stisihb}.}, then multiplied by a model for the stellar continuum in the vicinity of the absorption feature (2nd order polynomial), the parameters of which were allowed to vary during fitting. The final models for each feature were fit to the master spectrum using a Markov Chain Monte Carlo (MCMC) affine-invariant ensemble sampler \texttt{emcee} \citep{Foreman-Mackey2013}. Each species was fit separately, but all transitions listed in Table~\ref{table:lines} for an individual species were fit simultaneously. Lines arising from the same $E_l$ of each species were assumed to have the same $N_l$, and each velocity component was assumed to have the same $b$.

\begin{figure*}
\includegraphics[width=\textwidth]{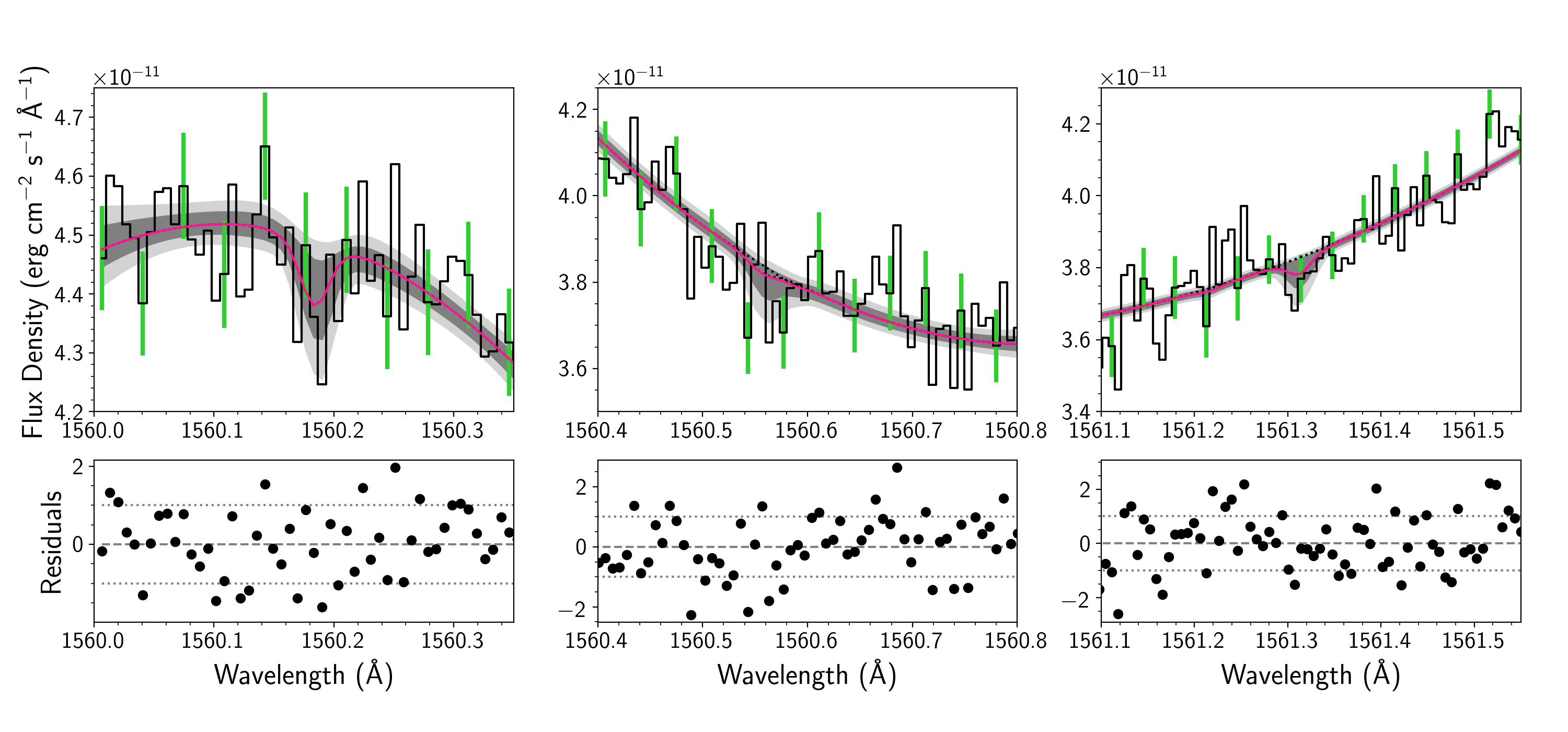}
\caption{The STIS spectra in the vicinity of the \CI\ absorption lines are shown in the top panels with black lines and 1-$\sigma$ error bars in green for every fifth data point for visual clarity. The best fit model (pink lines), individual velocity components (dotted black lines), 68\% confidence interval (dark grey shading), and 95\% confidence interval (light grey shading) are also shown. The model fitting was performed simultaneously for all lines shown. The bottom panels show the residuals of the best fit normalized by the data uncertainties.} \label{fig:CI_bestfit}
\end{figure*}

\begin{figure*}
\includegraphics[width=\textwidth]{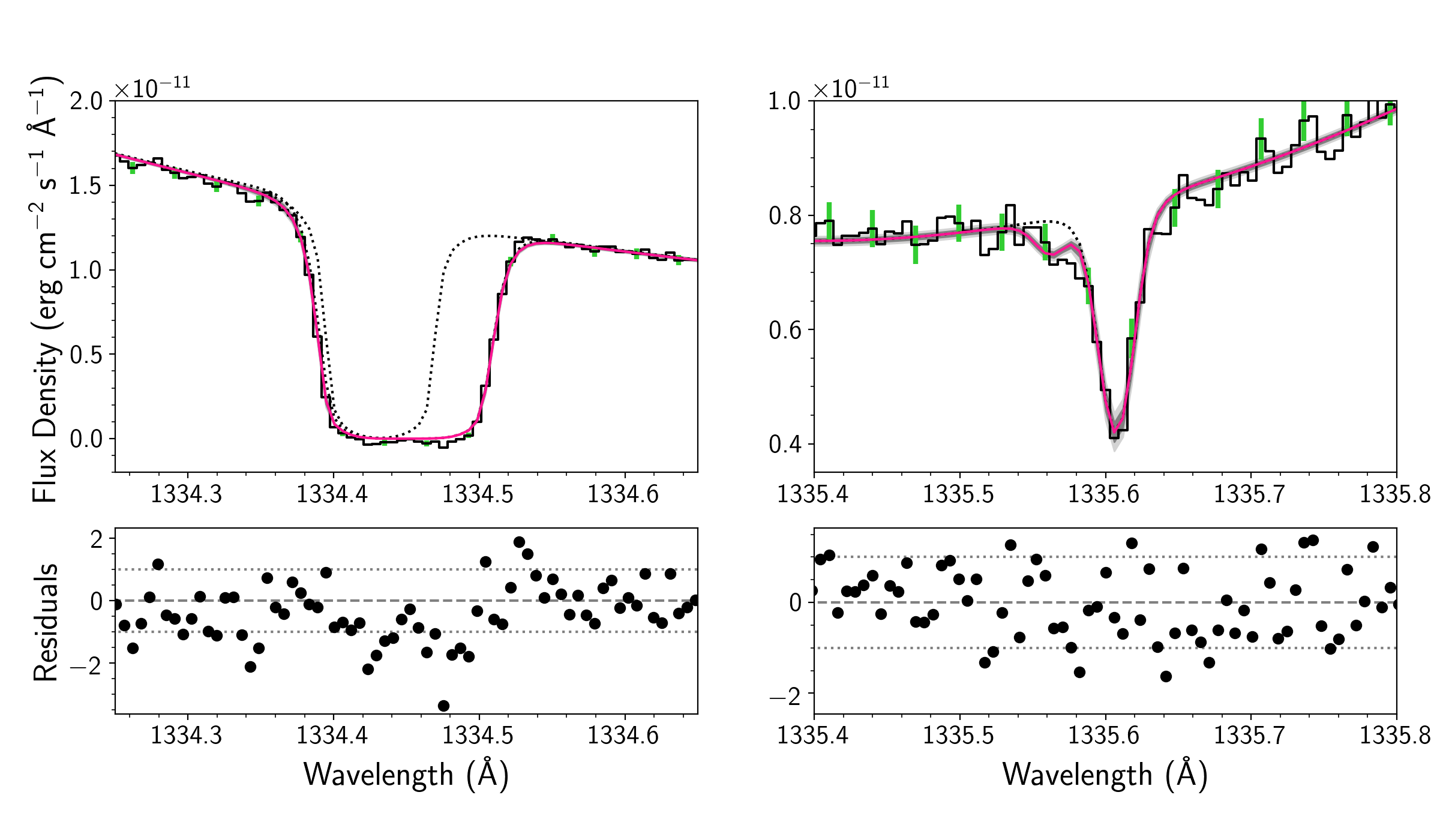}
\caption{Same as Figure~\ref{fig:CI_bestfit} for \CII.} \label{fig:CII_bestfit}
\end{figure*}

\begin{figure*}
\includegraphics[width=\textwidth]{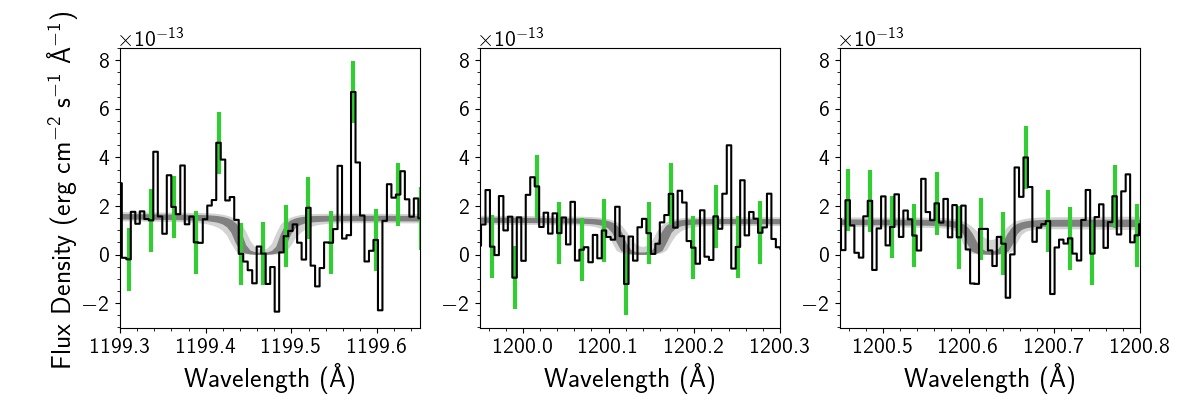}
\caption{Same as Figure~\ref{fig:CI_bestfit} for \NI. No best fit solution was found; $3 \sigma$ upper and lower limits are reported in Table~\ref{table:lines}.} \label{fig:NI_bestfit}
\end{figure*}

\begin{figure*}   
\includegraphics[width=\textwidth]{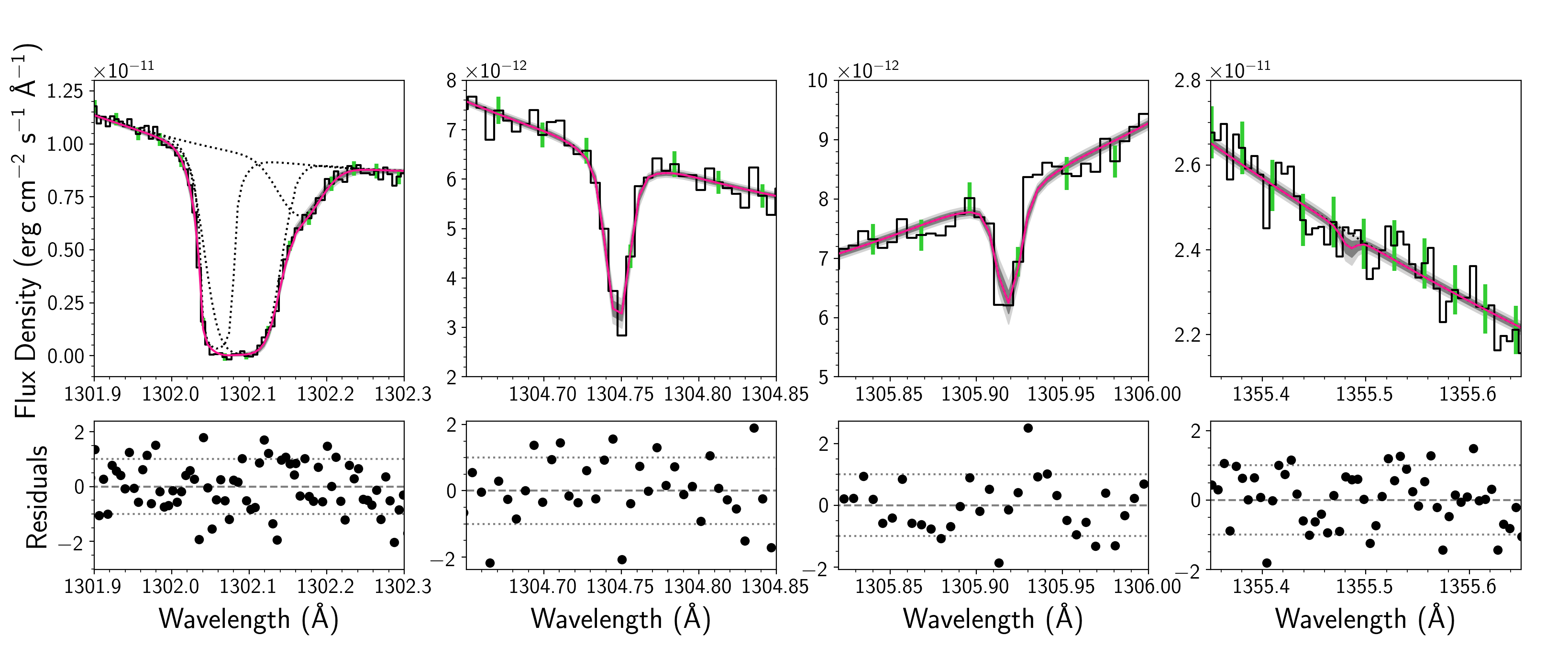}
\caption{Same as Figure~\ref{fig:CI_bestfit} for \OI. The third velocity component described in Section~\ref{subsec:OI_unknown} can be seen in the left panel at 1302.16 \AA.} \label{fig:OI_bestfit}
\end{figure*}

\begin{figure*}
\includegraphics[width=\textwidth]{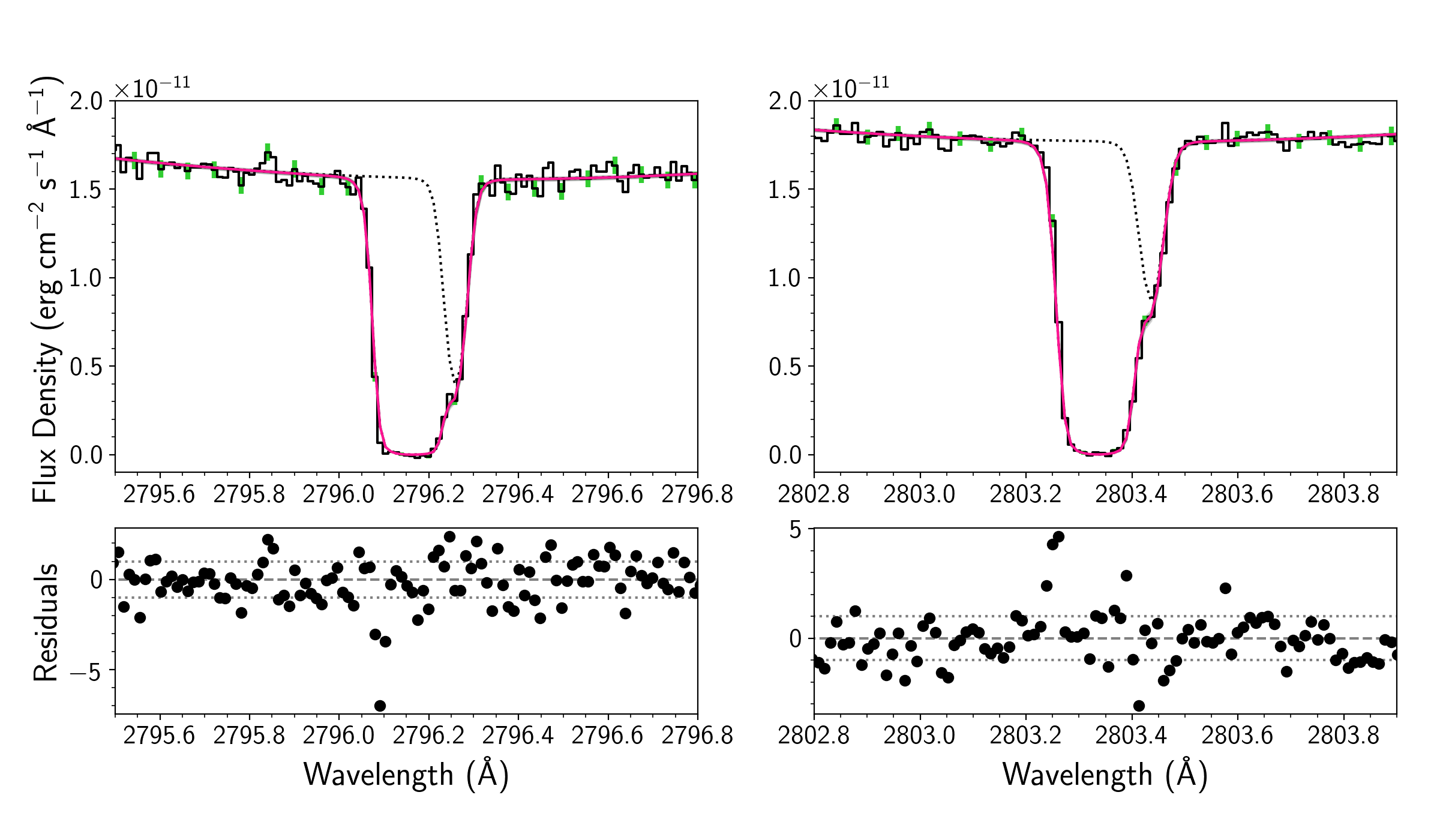}
\caption{Same as Figure~\ref{fig:CI_bestfit} for \MgII.} \label{fig:MgII_bestfit}
\end{figure*}

\begin{figure}
\includegraphics[width=0.47\textwidth]{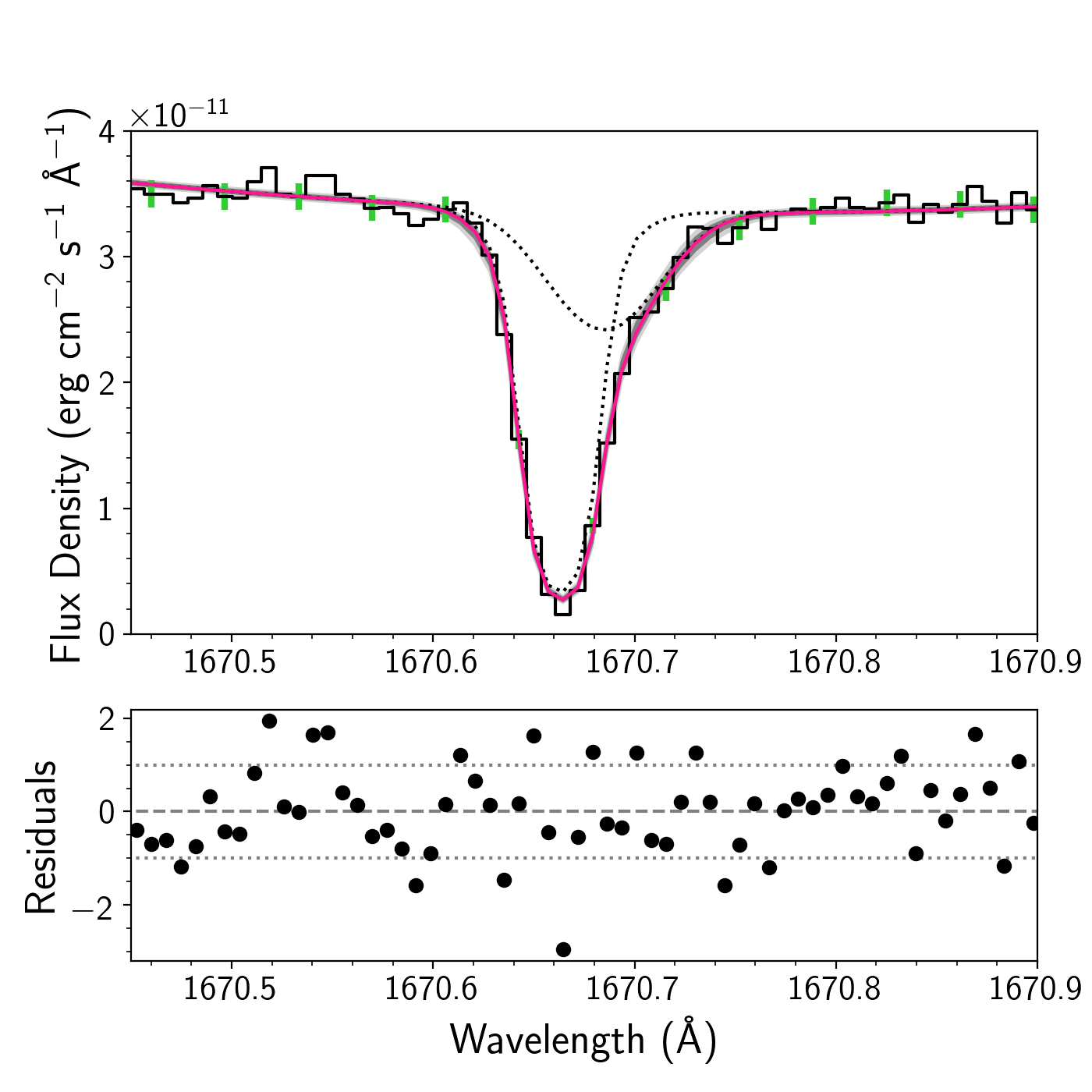}
\caption{Same as Figure~\ref{fig:CI_bestfit} for \AlII.} \label{fig:AlII_bestfit}
\end{figure}

\begin{figure*}
\includegraphics[width=\textwidth]{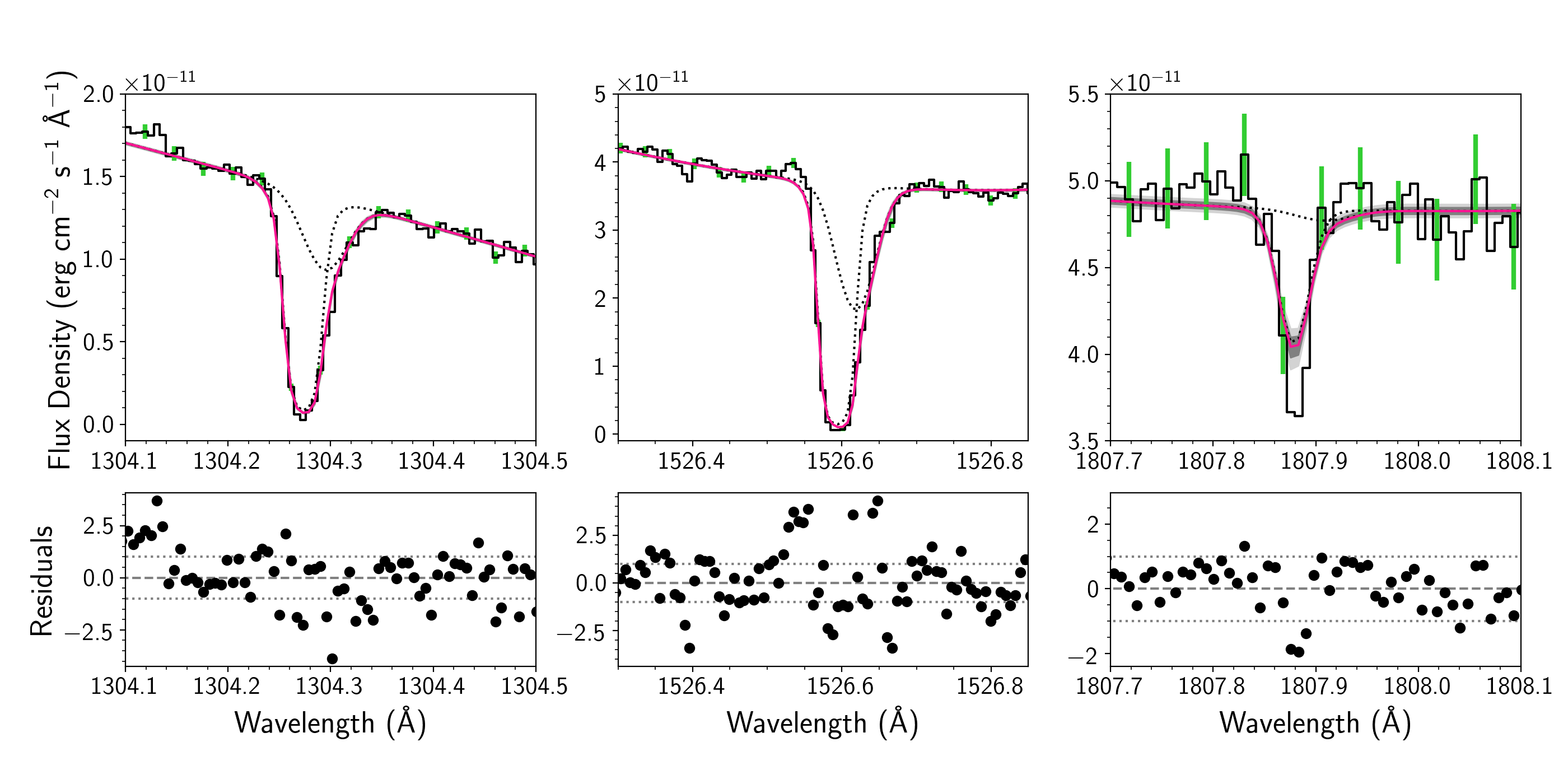}
\caption{Same as Figure~\ref{fig:CI_bestfit} for \SiII.} \label{fig:SiII_bestfit}
\end{figure*}

\begin{figure}
\includegraphics[width=0.49\textwidth]{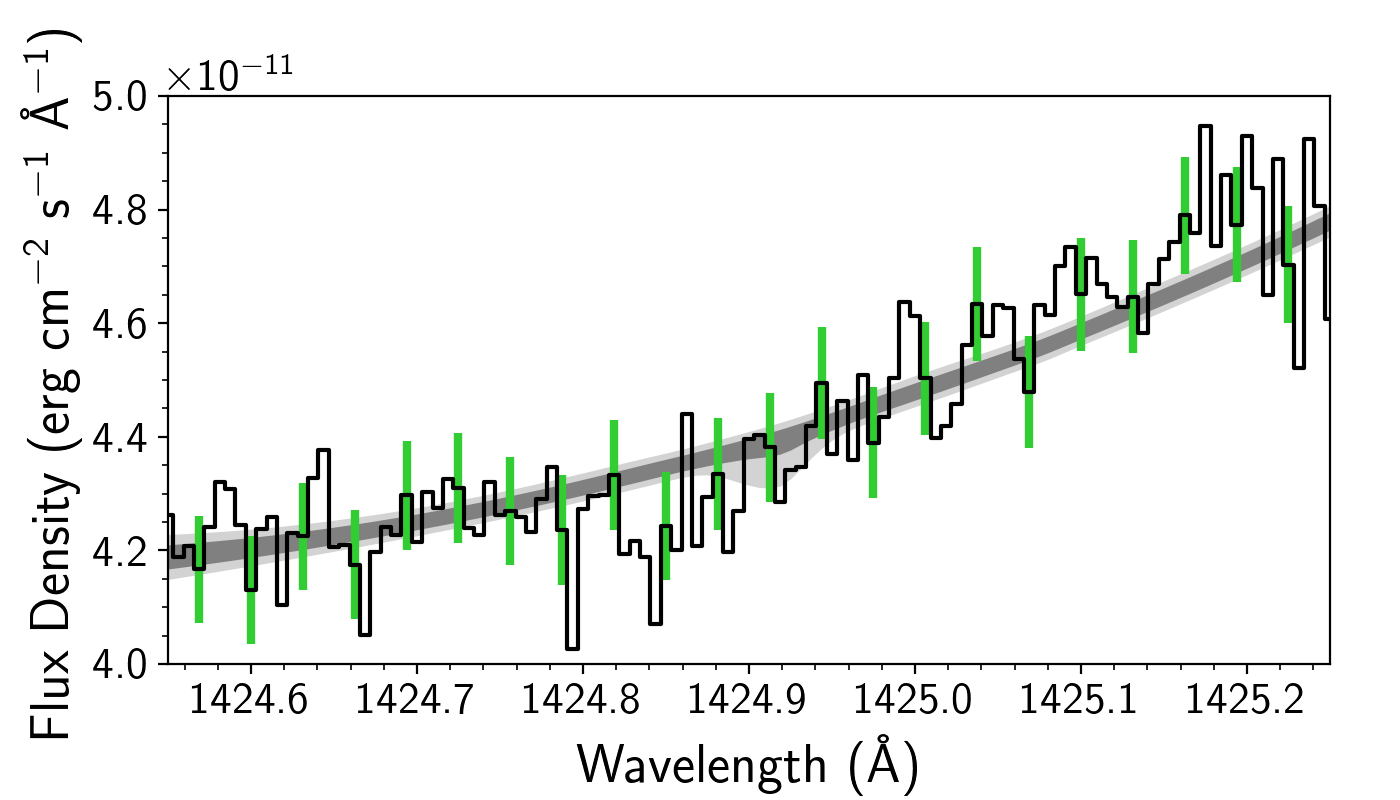}
\caption{Same as Figure~\ref{fig:CI_bestfit} for \SI. No best fit solution consistent with $N$(\SI) $> 0 \ \mathrm{cm}^{-2}$ was found; a $3 \sigma$ upper limit is reported in Table~\ref{table:lines}. } \label{fig:SI_bestfit}
\end{figure}

\begin{figure*}
\includegraphics[width=\textwidth]{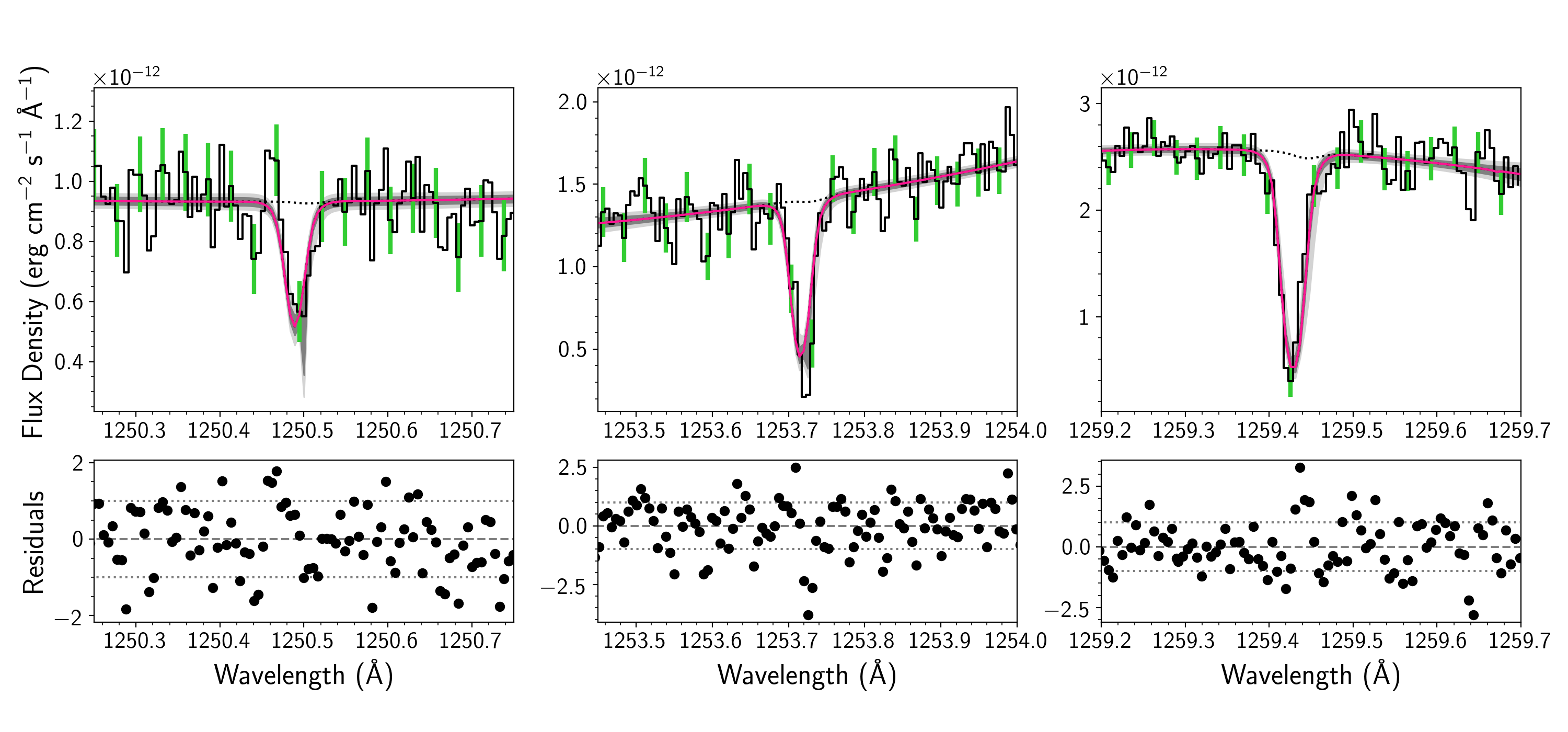}
\caption{Same as Figure~\ref{fig:CI_bestfit} for \SII.} \label{fig:SII_bestfit}
\end{figure*}

\begin{figure*}
\includegraphics[width=\textwidth]{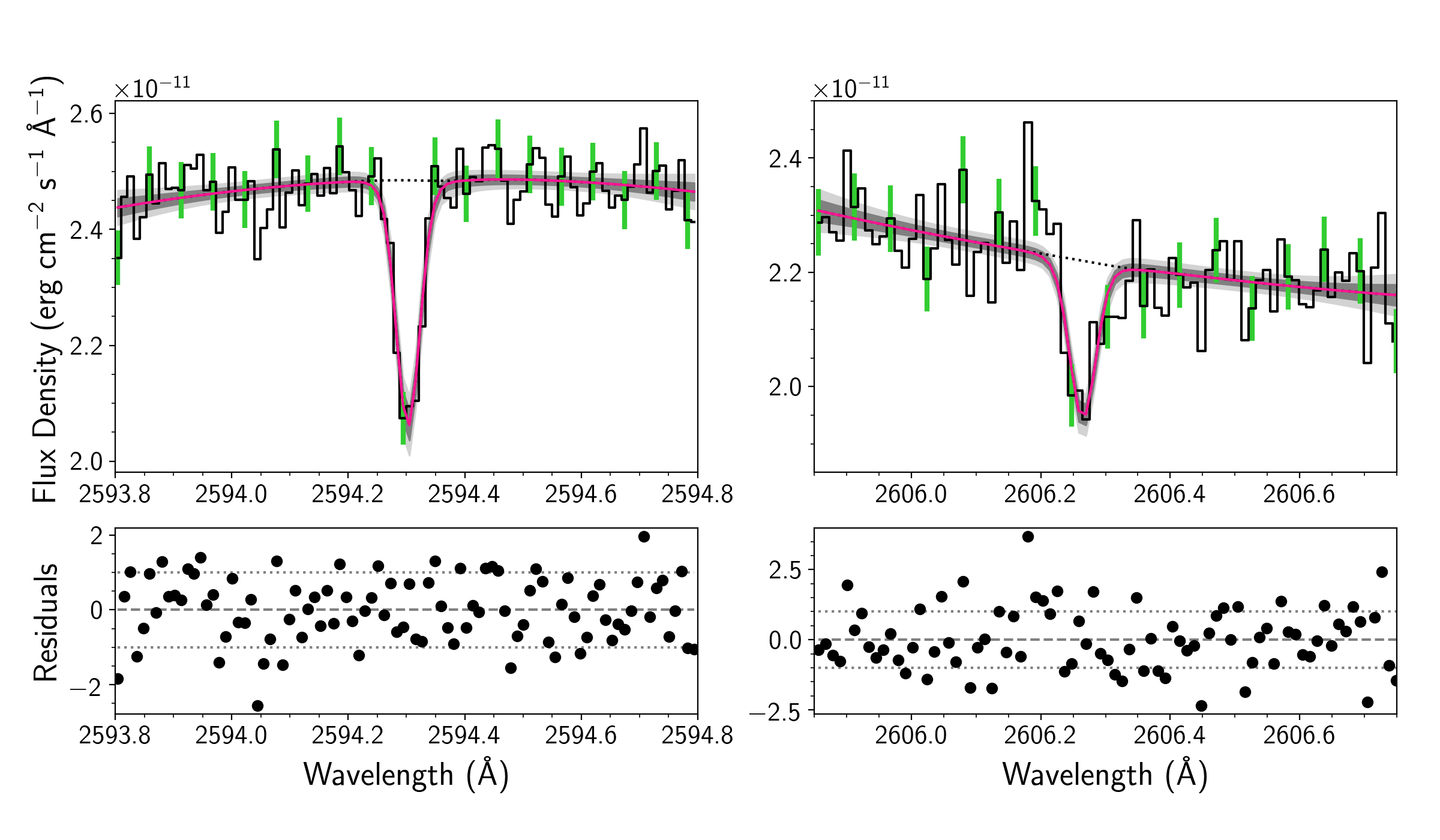}
\caption{Same as Figure~\ref{fig:CI_bestfit} for \MnII.} \label{fig:MnII_bestfit}
\end{figure*}

\begin{figure*}
\includegraphics[width=\textwidth]{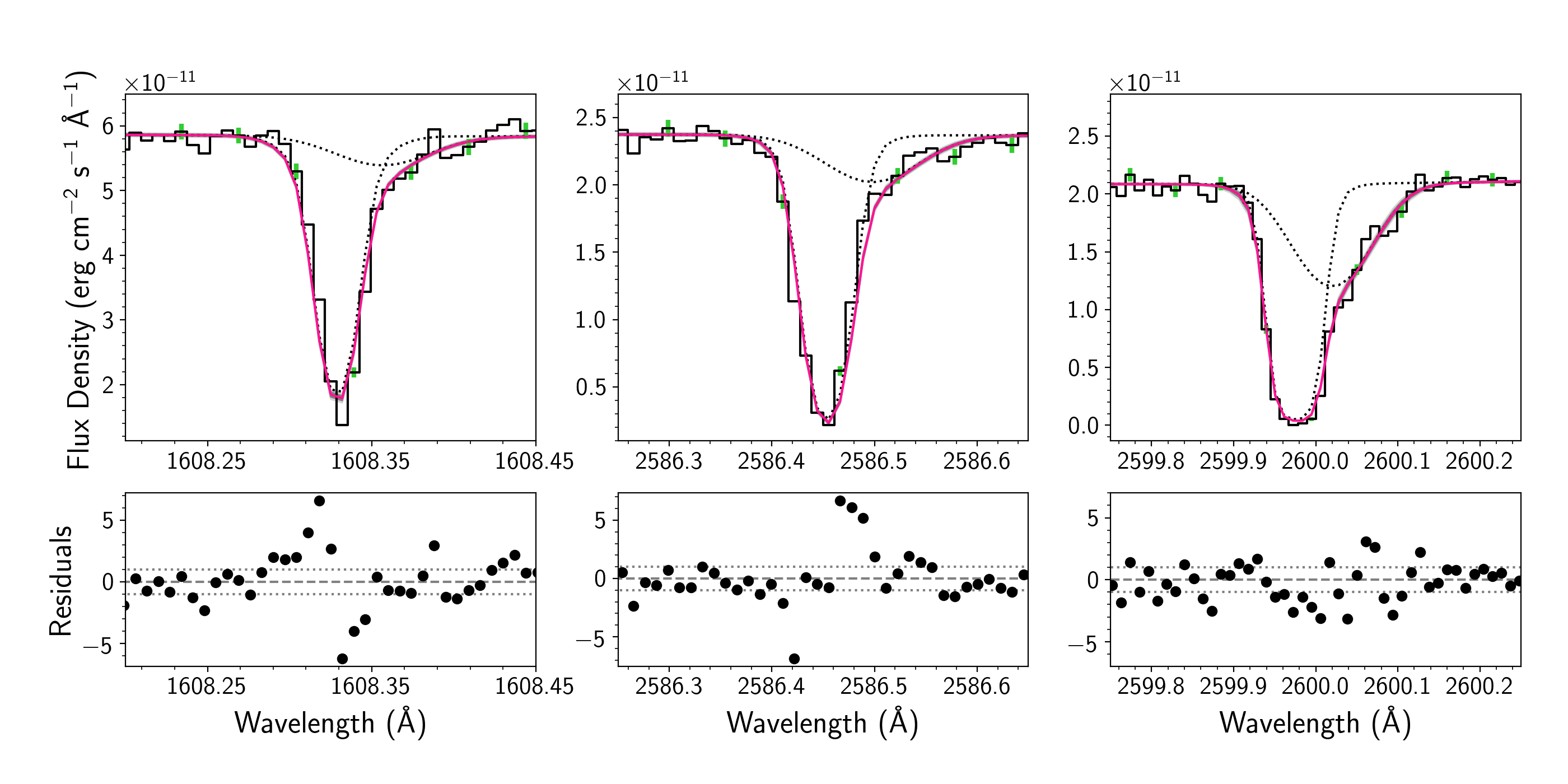}
\caption{Same as Figure~\ref{fig:CI_bestfit} for \FeII.} \label{fig:FeII_bestfit}
\end{figure*}

For the MCMC, we used $24-100$ walkers, depending on the number of free parameters, assumed a Gaussian likelihood and uniform priors for all parameters. We ran the chains for at least 50 autocorrelation times (10$^4-10^6$ steps) to approximate convergence, and removed an appropriate burn-in period based on the behavior of the walkers. Table~\ref{table:lines} lists the best fit (median) values and 68\% confidence intervals for the parameters $v_{\rm r}$, $\log_{10} N_l$, and $b$, derived from these parameters' marginalized posterior probability distributions. Note that in Table~\ref{table:lines} and throughout the paper, the radial velocities have the STIS absolute wavelength accuracy value (1.32 \kms) propagated into their confidence intervals unless otherwise stated. Figures \ref{fig:CI_bestfit} through \ref{fig:FeII_bestfit} show the data, best-fitting models, and residuals normalized by the data uncertainties. 

Lines arising from ground energy levels of species present in the local interstellar medium (ISM; \CII, \OI, \MgII, \AlII, \SiII, \SII, \MnII, and \FeII) show velocity components near $-23$~\kms\ and $-18$~\kms. 
Lines arising from species not abundant in the local ISM or from excited energy levels show a single velocity component near $-23$~\kms. 
We therefore identify velocity components near $-23$~\kms\ as originating in CS gas and components near $-18$~\kms\ as originating in IS gas (indicated in Table~\ref{table:lines}). This suggests that the \ion{Ca}{2} H \& K and \ion{Na}{1} D absorption detected at $-22$~\kms\ by \cite{Rebollido:2018} and \cite{Welsh2018} are circumstellar in origin. Two exceptions to this trend arose, which are further discussed in later sections: \OI~has a third absorption component at around $-1$~\kms\ (Section~\ref{subsec:OI_unknown}) and \MgII's redder velocity component (presumably IS in origin) appears at $-10$~\kms\ rather than $-18$~\kms\ (Section~\ref{subsection:ism_vel}).

We note that the MCMC fit finds a non-zero probability that N(\CII) = 0 cm$^{-2}$ for the CS component, but we present $\log_{10} N$(\CII) = 16.16$^{+0.18}_{-0.36}$ in Table~\ref{table:lines}. The shape of the marginalized distribution for the column density of \CII\ in the ground energy level consists of a strong, narrow peak at $\log_{10} N$(\CII) $\approx$ 16 and weak tail extending toward zero. This occurs because of degeneracy between the CS and IS absorbers in this heavily saturated line, which contains the bulk of the total reported C column density. Fortunately, the fit to the weaker \CII* absorption feature arising from the first excited fine structure level, which is uncontaminated by IS absorption, provides a strong constraint inconsistent with zero on $N$(\CII*). Since we expect the column density in the ground energy level to be greater than the first excited fine structure level, we restricted $\log_{10} N$(\CII) to be greater than 14, two orders of magnitude lower than the value at the strong narrow peak in the marginalized distribution. This prevented the $N$(\CII) confidence intervals from extending to arbitrarily small values and provides a better estimate of the error on the $N$(\CII) measurement. 

\SI\ was not significantly detected (Figure~\ref{fig:SI_bestfit}); therefore, a $3 \sigma$ upper limit on the CS column density was obtained assuming $v_{\rm r}$ between $-26$ and $-21$~\kms\ and $b$ between 0.1 and 10~\kms. \NI\ was marginally detected in the spectrum (Figure~\ref{fig:NI_bestfit}); however, we were unable to obtain a robust measurement and instead determined upper and lower limits for the \NI\ CS column density in the same way as for \SI. The signal-to-noise ratio of the data is low around 1200~\AA\ near the \NI\ lines and the lower limit on the column density should be treated with caution.  

For each species, Table~\ref{table:lines} also gives the total column density summing over all fine structure levels of the ground energy term. Note that \NI, \MgII, \AlII, \SII, and \MnII\ have only a single fine structure level in the ground term. For \SiII, \SI, and \FeII, we did not see significant absorption arising from the excited fine structure levels of the ground term. For these species, the total column density was assumed to be equal to the column density in the ground energy level. Since very small column densities could have been detected in the spectrum, this assumption is unlikely to affect any results.

\subsection{Additional \OI\ Absorption Component} \label{subsec:OI_unknown}

In the saturated \OI\ line arising from the ground energy level (1302.1685~\AA), we found a third velocity component at $v_{\rm r} = -0.76^{+2.25}_{-4.91}$~\kms\ not significantly detected in any other atomic absorption line (Figure~\ref{fig:OI_bestfit}). The column density in the third \OI\ component is smaller than the IS component's column density by a factor of 15-20, but has a similar $b$ value. Since time-variable velocity components redshifted with respect to the star are a sign of star-grazing exocomets, we examined the \OI\ feature in the two visits of STIS spectra taken $\sim3.5$ days apart, and find no significant variability near the velocity of the third component. Thus, we cannot confidently attribute this component to star-grazing exocomet activity. Neither can we rule out this explanation, as such activity may vary on timescales longer than the separation between the two visits.

\subsection{Carbon Monoxide with STIS and ALMA} \label{section:CO}

\begin{figure*}
\includegraphics[width=\textwidth]{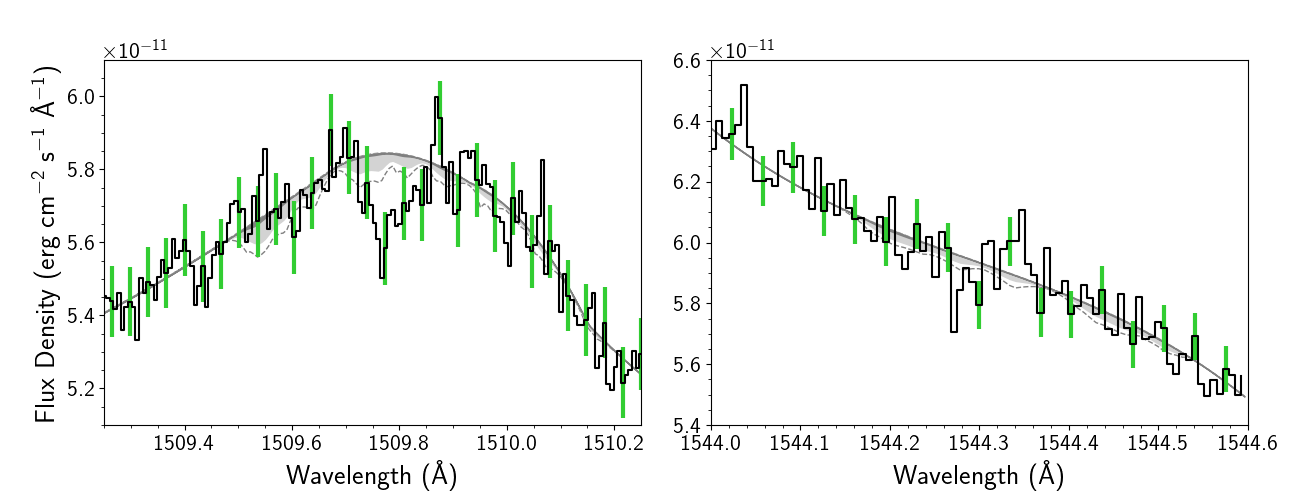}
\caption{Same as Figure~\ref{fig:CI_bestfit} for CO A-X (0-0) (\textit{left}) and A-X (1-0) (\textit{right}). No best fit solution consistent with $N(CO)_{total} > 0$ cm$^{-2}$ was found; the upper limit on the total CO column density from this fit assuming a Boltzmann distribution for the population of the rotational energy levels is reported in Table~\ref{table:lines}. The apparent absorption feature in the spectrum near 1509.8 \AA\ appears to be spurious. The 99.7\% confidence interval (dashed grey line) is shown to emphasize the low likelihood that real CO absorption is present in this spectrum.} \label{fig:CO_both_bestfit}
\end{figure*}

No CO gas absorption lines are seen in the STIS spectra, despite the availability of multiple strong electronic transition bands. The wavelength ranges around the $A-X \ (0-0)$ band at 1510~\AA\ and the $A-X \ (1-0)$ band at 1544~\AA\ are shown in Figure~\ref{fig:CO_both_bestfit}. The apparent absorption feature extending from 1509.7-1509.9 \AA\ is likely an instrument artifact despite its presence in all four spectra coadded to create the master spectrum in this region. If it were genuinely due to CO absorption, then signficant absorption would have been detected between 1509.5-1509.6 \AA\ as well. The narrow feature centered at 1509.77 \AA\ is present in only one of the four coadded spectra.

We calculated $3 \sigma$ upper limits on the line of sight CO column densities by producing models of the absorption bands similar to those for the atomic lines and fitting them to the data to determine confidence intervals. Upper limits were determined in two ways: 1.\ allowing the column densities in individual rotational energy levels ($J_l$) to independently vary and 2.\ assuming the population of the rotational energy levels follows a Boltzmann distribution. In both cases, only a single velocity component was included in the models. The free parameters in Case 1 were the CO column densities in the $J_l$ = 1--6 levels, the Doppler broadening parameter ($b$), and the radial velocity ($v_r$). The radial velocity was allowed to vary between $-26$ and $-21$~\kms\ and $b$ was allowed to vary between 0.1 and 5~\kms. For Case 2, the free parameters were excitation temperature ($T_{\rm ex}$), total CO column density, $b$, and radial velocity. The radial velocity and $b$ were allowed to vary over the same ranges as in Case 1, and $T_{\rm ex}$ was allowed to vary between 1 and 500~K. All upper limits on the CO column densities are reported in Table~\ref{table:lines}. 

\etaTel~was observed with ALMA during Cycle~2 on 2015 July 23 and September 17 (2013.1.01147.S). We executed the associated reduction scripts in \texttt{CASA} \cite[][version 4.4]{McMullin:2007} on the raw visibilities in order to regenerate the calibrated visibilities. The first scheduling block (SB) executed on 2015 July 23 showed significantly better quality than the second, with 44 antennas in the array spanning baselines of 15.1~m to 1.6~km and good weather. The second SB had only 34 antennas with baselines between 41.4~m and 2.1~km, and contributes little to the overall signal-to-noise of the ALMA observations. Between the two SBs, the total on-source integration time was 14~min.  

For these observations, the correlator was configured with one spectral window centered on the $^{12}$CO $J=2-1$ line at 230.5~GHz and a second centered at 220.4~GHz, both with 3840 channels covering 58593.8~kHz total bandwidth for a spectral resolution of 30.52~kHz. The remaining two spectral windows, centered at 218.5 and 233.0~GHz, were intended to detect only continuum emission at low spectral resolution (128 channels covering 2~GHz). All deconvolution and imaging was performed using the \texttt{clean} task in \texttt{CASA} (version 4.3.1).

Neither continuum emission nor $^{12}$CO line emission from \etaTel\ are detected in these ALMA data. With natural weighting, the continuum RMS is $0.051$~mJy~beam$^{-1}$ for a beam size of $0\farcs29\times0\farcs22$. Assuming the \etaTel\ disk geometry inferred from imaging observations with T-ReCS on Gemini South at 11.7 and 18.3~$\mu$m \citep{Smith2009}, we estimate that the disk emission would be spread over $\sim7$ beams given the angular resolution of these  ALMA observations. Thus, the non-detection of continuum emission implies a $3 \sigma$ upper limit on the disk continuum flux of $<1.1$~mJy at 1.33~mm.  We can use this upper limit to estimate an upper limit on the total dust mass in the disk using $M_{dust} = F_\nu d^2/(\kappa_\nu B_\nu(T_{dust}))$, where $F_\nu$ is the millimeter flux density, $d$ is the distance, $\kappa_\nu$ is the dust opacity, and $B_\nu$ is the Planck function at the dust temperature, $T_{dust}$.  We use a dust opacity of $\kappa_\nu = 2.3$~cm$^2$~g$^{-1}$ as is common in the literature \citep{Beckwith1990}, but note that the significant uncertainty on this value dominates the uncertainty in the resulting dust mass.  \cite{Pollack1994} obtain dust opacities that are several orders of magnitude smaller ($10^{-2}-10^{-3}$~cm$^2$~g$^{-1}$), which would increase the calculated dust masses considerably.  Assuming a dust temperature of 100~K (typical for a warm inner dust belt), yields a total dust mass of less than $7.22\times10^{25}$~g or 0.012~$M_\oplus$. This is consistent with the $T_{dust}$=162 K estimate from \cite{RiviereMarichalar2014} based on \textit{Herschel} [\CII] data. For a colder outer dust belt with a temperature of 40~K, the total dust mass would be less than $1.96\times10^{26}$~g or 0.033~$M_\oplus$.

The RMS for the $^{12}$CO line is $11$~mJy~beam$^{-1}$ measured in channels with width $0.2$~km~s$^{-1}$.  Making the same assumptions about the disk geometry as before, this yields a $3\sigma$ upper limit on the integrated line intensity of $0.23$~Jy~km~s$^{-1} = 1.7\times10^{-21}$~W~m$^{-2}$. This upper limit is high compared to $^{12}$CO gas emission measured with ALMA for other A-star debris disks \cite[e.g., 49 Cet and HD~32297;][]{Hughes2017, MacGregor2018}. Since the inclination of the \etaTel\ disk is only constrained to be within $20^\circ$ of edge-on \citep[$i > 70^\circ$;][]{Smith2009}, the sensitive upper limits from STIS on the line-of-sight CO column densities in the various rotational levels cannot reliably be used to estimate upper limits on pure rotational CO emission. Future targeted ALMA observations with lower resolution and deeper sensitivity limits will be able to detect smaller CO gas masses or place much more stringent upper limits.

\subsection{Mystery absorber at 1594 \AA} \label{subsec:Mystery}

\begin{figure*}
\begin{center}
\includegraphics[width=0.75\textwidth]{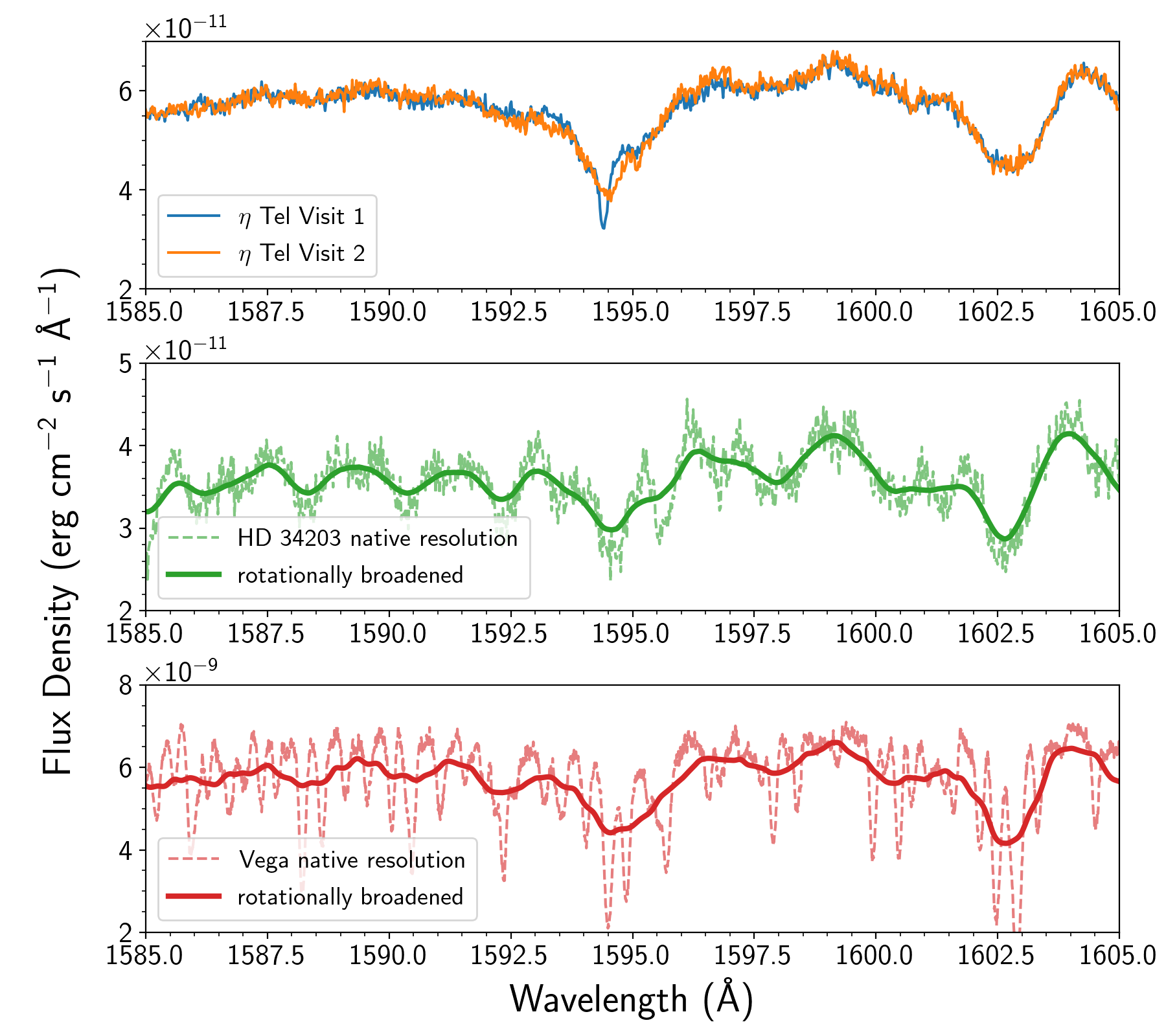}
\caption{\textit{Top panel:} Comparison of the unidentified variable absorption line at 1594 \AA\ in the two visits of STIS data taken $\sim$3.5 days apart. The data have been binned by a factor of three for clarity, and the Visit~2 fluxes shifted down by 5$\times$10$^{-12}$ erg cm$^{-2}$ s$^{-1}$ \AA$^{-1}$ to better match the Visit~1 fluxes (see Section~\ref{sec:ObservationsReductions}). \textit{Middle panel:} An archival STIS E140H spectrum of HD~34203 (18 Ori; A0V) is shown at the native spectral resolution (dashed, light colored line) and rotationally broadened (solid line) to match $\eta$~Tel's $v \sin i$. \textit{Bottom panel:} Archival STIS E140H spectrum of Vega (A0V) shown at the native spectral resolution (dashed, light colored line) and rotationally broadened (solid line) to match $\eta$~Tel's $v \sin i$. The unidentified absorption feature in the \etaTel\ spectra is narrower than other rotationally broadened stellar features and does not appear in the two comparison spectra of A0V stars without edge-on CS disks. We conclude that the mystery feature in the \etaTel\ spectra is unlikely to arise in the star.} \label{fig:mystery_line}
\end{center}
\end{figure*}

We identify an absorption line at $\sim1594$~\AA\ originating from an unknown species, which was also observed in STIS spectra of the 49~Cet debris disk \citep{Miles2016}. In the 49~Cet spectra, the absorption was variable on the timescale of a few days, suggesting it originated from star-grazing comets. Similarly, the feature in the \etaTel\ data shows strong variability between the two visits (top panel of Figure~\ref{fig:mystery_line}), with a narrow absorption feature at 1594.4~\AA\ disappearing between the first and second visits. There also appear to be shallower transient features on either side of the narrow feature. 

During our attempts to identify the species responsible for this absorption feature, multiple databases of atomic and ionic transitions were searched. Unfortunately, the feature shows no clear stable, unvarying absorption component that could help identify the rest wavelength, as was also the case for 49~Cet. The most promising candidate species (e.g., \ion{O}{2}, \ion{Si}{1}, \ion{Fe}{2}) produce additional absorption lines within the wavelength range of the STIS spectra that are not seen or do not vary between visits. Therefore, we found no solid identification for the mystery species.

Some absorption lines in the 49~Cet spectra (\CII\ at 1335~\AA\ and \ion{C}{4} at 1550~\AA) showed variations characteristic of star-grazing comets, making it easier to associate the mystery feature with such activity. This explanation is more difficult to accept in the case of \etaTel, as no other variable features are seen in the spectra. 

The deep absorption feature at 1594.4~\AA\ in the Visit~1 spectrum is narrower than other stellar absorption lines in the \etaTel\ spectra, suggesting that the mystery feature does not arise in the atmosphere of this rapidly rotating star. To demonstrate this, we examined archival STIS spectra of early A stars without CS gas along the line of sight. The lower two panels in Figure~\ref{fig:mystery_line} show the STIS spectra of HD~34203 (18 Ori), an A0V star not known to host a CS disk, and Vega (A0V), an A0V star hosting a well-known face-on debris disk. We convolved the HD~34203 and Vega spectra with a rotational kernel to visually match the rotationally-broadened features of \etaTel, which is viewed edge-on and therefore has maximal rotational broadening. We find no evidence of a similar narrow feature at 1594~\AA\ in either spectrum, additionally indicating that it is not stellar in origin. 

\section{Velocity structure of the line of sight} \label{section:velocity}

\subsection{Radial velocity of the ISM} \label{subsection:ism_vel}

The \etaTel\ STIS spectra provide new measurements of the local ISM in the direction of \etaTel, which we compare to a kinematic model of the ISM\footnote{\url{http://lism.wesleyan.edu/LISMdynamics.html}} \citep{Redfield2008}. According to the model, the \etaTel~line of sight ($l=342.90^{\circ}$, $b=-26.21^{\circ}$, $d=47.36$~pc; \citealt{Brown2018}) does not traverse any known ISM clouds. However, the boundaries of individual local clouds are not well sampled by current data sets, and the \etaTel\ sight line probably does traverse at least one cloud. The kinematic model indicates \etaTel's sight line passes near the LIC, G, Aql, and Vel clouds, which have radial velocities of $-16.62 \pm 1.17$~\kms, $-18.51 \pm 1.51$~\kms, $-10.28 \pm 1.03$~\kms, and $-27.59 \pm 1.37$~\kms, respectively. The radial velocity of the absorption components identified as IS in Table~\ref{table:lines} have a weighted average radial velocity $v_{\rm r} = -17.8\pm0.7$~\kms\ (excluding \MgII), which is consistent with the LIC and G cloud velocities. We conclude that these local IS clouds have a somewhat greater extent than previously thought. 

The \MgII\ absorption lines are not well fit with an IS component near $-18$~\kms, but rather the IS component appears at $-10.16^{+1.31}_{-1.34}$~\kms. We created alternate models of the \MgII\ lines that contained different numbers of absorption components and fit them using various parameter restrictions to see if a model with a component at the regular CS velocity ($-20$ to $-25$~\kms) and a component at the regular IS velocity ($-20$ to $-15$~\kms) was consistent with the data. However, these fits were poorly constrained and an absorption component near $-10$~\kms\ was still required. An IS absorber at $-10$~\kms\ is consistent with the radial velocity of the Aql cloud. It appears that the \MgII\ IS absorption component arising in the LIC/G clouds is lost in the saturated CS \MgII\ absorption component, but absorption arising in the Aql cloud is detectable for this feature alone due to the high oscillator strengths of the \MgII\ transitions.

\subsection{Radial velocity of the star} \label{subsection:star_vel}

\begin{figure}
\includegraphics[width=0.48\textwidth]{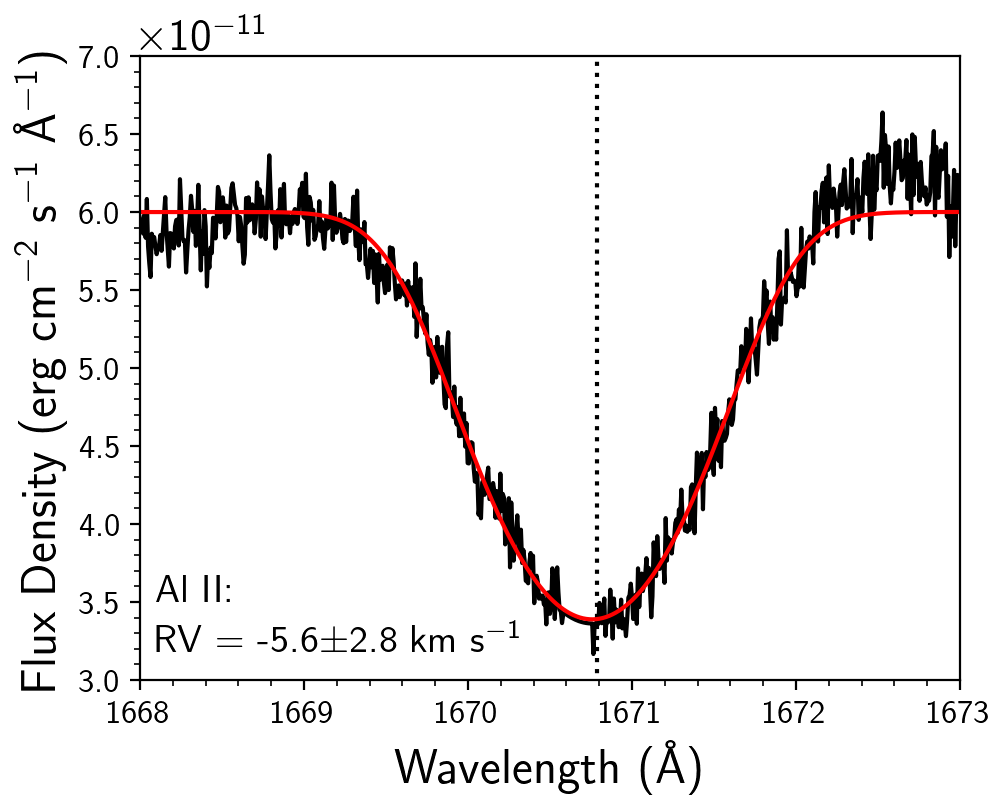}
\caption{The \AlII\ photospheric absorption line in the STIS spectrum (black line) and the best fitting rotationally broaded Gaussian model for the photospheric absorption (red line) are shown. The wavelength range including IS and CS absorption features was excised and the gap interpolated over with a 2nd order polynomial before fitting. The vertical dotted line shows the position of the transition's rest wavelength. The stellar radial velocity corresponding to the best fitting model is $-5.6\pm$2.8~\kms.} \label{fig:RV}
\end{figure}

The radial velocity of the central star is critical for interpreting the kinematics of the CS absorbers. \etaTel's radial velocity is reported in the literature as $v_\star = +13.0 \pm 3.7$~\kms\ \citep{Kharchenko2007}, $-3.0\pm 3.0$~\kms\ \citep{Rebollido:2018}, and $-5.0 \pm 1.5$~\kms\ \citep{Rebollido:2020}. Measuring the radial velocity of a rapidly rotating young A star is notoriously difficult given the large width and blending of the photospheric lines. In order to clearly determine the stellar radial velocity, we fitted a rotationally broadened Gaussian model to a clean, single stellar photospheric absorption line in the STIS data (\AlII\ at 1670 \AA; Figure~\ref{fig:RV}). We allowed all model parameters ($v_\star$, FWHM, $v \sin i$, and linear limb darkening) to vary within a range of reasonable values, and we find $v_\star = -5.6 \pm 2.8$~\kms\ relative to the rest wavelength of \AlII. Our measured radial velocity is consistent with \cite{Rebollido:2018} and \cite{Rebollido:2020}. In the rest of this paper, we adopt the $v_\star$ value from our \AlII\ analysis ($-5.6$~\kms) as the stellar radial velocity.

\section{Elemental Abundances} \label{section:abundances}

\subsection{Ionization Balance in the CS Gas} \label{subsec:ionization}

To determine the elemental abundances in the CS gas, we must consider how much gas is present in all the dominant ionization states. Ideally, the column density in every ionization state would be measured and summed to provide a total elemental column density; however, the necessary transitions are not all present in the wavelength range of our data. Therefore, we make use of ionization balance calculations for the circumstellar environment of an A0V star. The neutral fractions (${n_0}/{n_+}$) from \citet{Fernandez:2006} appear in Table~\ref{table:abundances}. Using these fractions, we can estimate the abundance of an element in an unseen ionization state from measurements of the column density in a different ionization state. 

\begin{deluxetable*}{lCLcRR}
\tablecolumns{2}
\tablewidth{0pt}
\tablecaption{Absolute and Relative Circumstellar Elemental Abundances \label{table:abundances}} 
\tablehead{\colhead{Element} & \colhead{${n_0}/{n_+}$ \tablenotemark{a}} & 
	\colhead{$\log_{10} N_{\rm total}$} & \colhead{Element} & \colhead{$\log_{10} \mathrm{ratio}$} & \colhead{$\log_{10}$~ratio \tablenotemark{b}} \\
    \colhead{} & \colhead{} & \colhead{$\mathrm{cm}^{-2}$} & \colhead{Pair} & \colhead{\etaTel} 
    & \colhead{solar}}
\startdata
C & 0.01 & $16.17^{+0.18}_{-0.36}$ \ \tablenotemark{c} & C/Fe & $2.69^{+0.18}_{-0.36}$ 
& $0.94 \pm 0.09$ \\
N & $\sim 1$ \ \tablenotemark{d} & \multicolumn{1}{c}{$13.53 - 17.24$} & N/Fe & $0.14 - 3.69$ 
& $0.38 \pm 0.14$ \\
O & $\sim 1$ \ \tablenotemark{d} & $16.17^{+0.21}_{-0.30}$ & O/Fe & $2.69^{+0.21}_{-0.29}$ 
& $1.24 \pm 0.09$ \\
Mg & $6 \times 10^{-6}$ & $14.16^{+0.08}_{-0.07}$ & Mg/Fe & 0.69$^{+0.08}_{-0.07}$ & $0.10 \pm 0.08$ \\
Al & $1 \times 10^{-7}$ & $12.44^{+0.04}_{-0.02}$ & Al/Fe & $-1.03^{+0.03}_{-0.03}$ 
& $-0.98 \pm 0.11$ \\
Si & $2 \times 10^{-7}$ & $14.04^{+0.03}_{-0.03}$ & Si/Fe & $0.57^{+0.03}_{-0.03}$ 
& $0.09 \pm 0.09$ \\
S & $5 \times 10^{-4}$ & $14.35^{+0.03}_{-0.06}$ & S/Fe & $0.87^{+0.04}_{-0.05}$ 
& $-0.25 \pm 0.09$ \\
Mn & $\sim 6 \times 10^{-6}$ \ \tablenotemark{d} & $11.76^{+0.03}_{-0.03}$ & Mn/Fe & $-1.72^{+0.03}_{-0.03}$ & $-1.95 \pm 0.09$ \\
Fe & $3 \times 10^{-6}$ & $13.48^{+0.01}_{-0.01}$ & C/O & $-0.03^{+0.35}_{-0.40}$ 
& $-0.30 \pm 0.06$ \\
\enddata
\tablenotetext{a}{Calculated ionization fractions for CS gas around an A0V star from \citet{Fernandez:2006}. $n_e$=100 cm$^{-2}$ and $d$=100 au are assumed.}
\tablenotetext{b}{From \cite{Lodders2003}.}
\tablenotetext{c}{Total C column density measured via direct observation of \CI\ and \CII\ in the STIS data.}
\tablenotetext{d}{Assumed from gas ionization balance; see main text.}
\end{deluxetable*}

Calculated neutral fractions for N, O, and Mn do not appear in \citet{Fernandez:2006}. However, we inferred the neutral fractions for these species given their ionization energies. The first ionization energies of N and O (14.53~eV and 13.62~eV) are greater that of H (13.60 eV). Therefore, the bulk of the N and O should be in their neutral states (\NI\ and \OI), and we adopted the measured \NI\ and \OI\ column densities as the total N and O column densities. The first ionization energy of Mn (7.43~eV) is similar to that of Mg (7.65~eV). Therefore, we adopted the calculated Mg neutral fraction value from \citet{Fernandez:2006} for Mn. 

The neutral fractions show that the \etaTel\ CS gas is highly ionized, with only trace amounts of neutral Mg, Al, Si, S, Mn, and Fe. Our measurements of the CS column densities of \CI\ and \CII\ in all fine structure energy levels of the ground term provide a check on the ionization balance calculations. The total circumstellar \CI\ column density in all fine structure energy levels is $9.43 \times 10^{11} \ \mathrm{cm}^{-2}$ and the total circumstellar \CII\ column density in all fine structure levels is $1.48 \times 10^{16} \ \mathrm{cm}^{-2}$, giving a neutral fraction of $6 \times 10^{-5}$. This is smaller than the calculated ionization balance of 0.01 from \citet{Fernandez:2006}, which was estimated assuming $n_e$=100 cm$^{-3}$ and a 100 au distance from the star, indicating that the \etaTel\ gas is closer to the star and/or less dense. However, this does not significantly affect our general conclusions about the relative elemental abundances in the CS gas. 

\subsection{Total Elemental Column Densities} \label{subsec:tot_cols}

Using this information about the dominant ionization state for each element, we determined the total CS column densities for all studied elements, given in Table~\ref{table:abundances}. Since Mg, Al, Si, S, Mn, and Fe should be primarily in the first ionized state, the total column densities for these elements were taken to be equal to the total CS column densities of \MgII, \AlII, \SiII, \SII, \MnII, and \FeII\ in all fine structure energy levels of the ground term (given in Table~\ref{table:lines}). Nitrogen and oxygen, on the other hand, should be primarily neutral; therefore, the total column densities for these elements were taken to be equal to the total CS column densities of \NI\ and \OI\ in all fine structure levels of the ground term. For carbon, the total elemental CS column density is the sum of the measured CS column densities in all fine structure levels for both the neutral and first ionized states. 

Because the marginalized probability distributions for the fitted column densities are often not Gaussian, we determined the error bars for the total elemental column densities using a Monte Carlo routine. Using C as an example, 10$^6$ random draws were made from each of the probability distributions of $N$(\CI), $N$(\CI*), $N$(\CI**), $N$(\CII), and $N$(\CII*). The error bars for $N($C$)_{\mathrm{total}}$ were then taken as the 68\% confidence interval of the distribution of the summed column densities.

\subsection{Relative Elemental Abundances}

\begin{figure*}
\includegraphics[width=\textwidth]{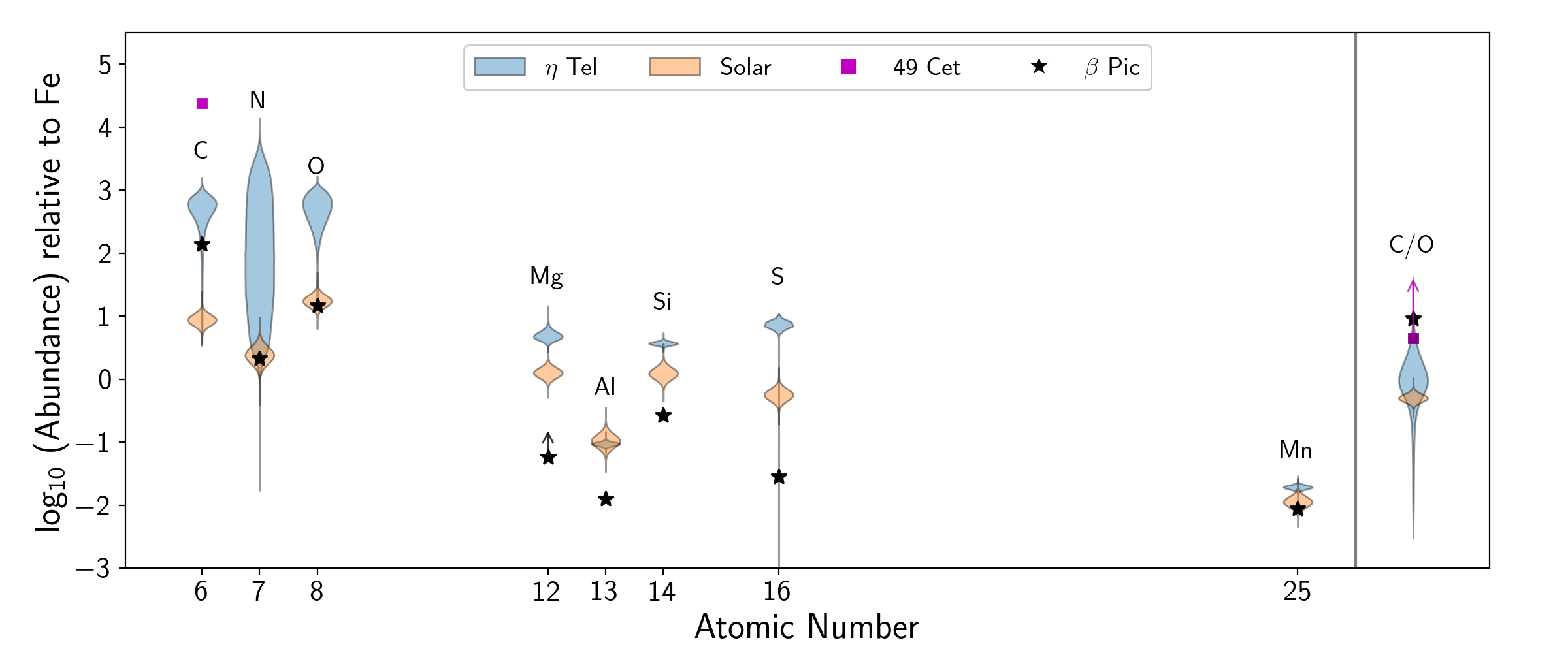}
\caption{Abundances in the \etaTel~gas disk (blue) from this work compared to solar abundances (orange) from \citet{Lodders2003} as a function of atomic mass. Note that the C/O abundances on the far right do not correspond to an atomic mass on the horizontal axis. The violin plot illustrates the kernel probability density for each measurement, i.e., the width of the shaded area represents the proportion of models located at that abundance value. Values for the 
	49~Cet and $\beta$~Pic abundances obtained with UV absorption spectroscopy are plotted with purple square and black star symbols, respectively.  The 49~Cet abundances are from \citet{Roberge2014}. The $\beta$~Pic N/Fe abundance is from \citet{Wilson2019}; the other $\beta$~Pic abundances are from \citet{Roberge2006}. The arrow indicates that the $\beta$~Pic Mg/Fe abundance is a lower limit. Not shown is the log O/Fe = 3.6-5.4 range for $\beta$~Pic from \citet{Brandeker2016}, based upon far-IR emission measurements from \emph{Herschel}.} \label{fig:abundances}
\end{figure*}

A similar Monte Carlo process was used to calculate the relative abundances C/Fe, N/Fe, O/Fe, Mg/Fe, Al/Fe, Si/Fe, S/Fe, Mn/Fe, and C/O in the \etaTel\ CS gas. 10$^6$ random draws were made from the marginalized probability distributions representing the total C, N, O, Mg, Al, Si, S, Mn, and Fe column densities to construct a probability distribution for each ratio. We report the best-fit (median) values and 68\% confidence intervals in Table~\ref{table:abundances}, along with the solar relative elemental ratios from \citet{Lodders2003}. Figure~\ref{fig:abundances} shows our relative abundance measurements compared to solar. In the \etaTel\ CS gas, C, O, Mg, Si, S, and Mn all show super-solar abundances relative to Fe. On the other hand, the Al/Fe ratio is consistent with solar abundance. While N may also show super-solar abundance relative to Fe, the large limits on the \NI\ column density allow the N/Fe ratio to be consistent with the solar value. 

Given the critical role that C and O play in the formation of planetary material and the debate about the C/O ratio in the $\beta$~Pic CS gas \citep[see][and references therein]{Brandeker2016}, we took a close look at this ratio for \etaTel. We derived the C/O ratio and 68\% confidence interval ($\log_{10} \mathrm{C/O} = -0.03^{+0.35}_{-0.40}$) in the same manner described above. In contrast to $\beta$~Pic and 49~Cet, which show super-solar abundance of C relative to O in UV absorption spectra \citep{Roberge2006,Roberge2014}, we find that the C/O ratio in the \etaTel\ CS gas is consistent with the solar C/O ratio. The final picture of the \etaTel\ CS gas is one that is enriched in lighter elements relative to iron, but does not show an abundance anomaly among the volatile elements. The fact that Al alone among the siderophile elements shows solar abundance relative to Fe appears puzzling, but may point to the mineralogy of the \etaTel\ CS dust.

\section{Gas Dynamics} \label{section:dynamics}

\subsection{Radiation Pressure} \label{sub:rad}

To analyze the dynamics of the CS absorbers, we computed the ratio ($\beta$) of radiation pressure to gravitational force that CS atoms and molecules receive from the \etaTel\ central star. Particles experiencing $\beta > 0.5$ will be pushed onto hyperbolic orbits and ejected from the system. To compute the stellar radiation pressure, we modeled the flux from the star using a synthetic spectrum from the Castelli-Kurucz atlas\footnote{\url{https://www.stsci.edu/hst/instrumentation/reference-data-for-calibration-and-tools/astronomical-catalogs/castelli-and-kurucz-atlas}} \citep{Castelli2003} with solar metallicity and $T_\mathrm{eff} = 9750$~K. Assuming a stellar radius of 1.61~R$_{\odot}$ \citep{Rhee2007} and distance to \etaTel\ of 47.36~pc \citep{GaiaDR2}, we find that the synthetic spectrum matches the STIS spectrum within 50\% in the $>$1600 \AA\ spectral regions where they overlap. However, at $<$1600 \AA, the agreement is poorer with the synthetic spectrum deviating from the observed spectrum by factors up to 3$\times$.

For each species listed in Table~\ref{table:lines}, we include transitions with $E_l = 0$~cm$^{-1}$ using the line lists from \citet{Kurucz2011}\footnote{\url{http://kurucz.harvard.edu/linelists/gfall}}, computing the total radiation pressure in the same manner as \cite{Beust1993}:

\begin{equation} \label{eq:radiationpressure}
F_\mathrm{rad} = \frac{1}{4\pi \epsilon_0}\frac{\pi e^2}{m_e c^2 } \sum_{i=1} f_i F_{\nu,i},
\end{equation}

\noindent where $e$ is the elementary charge, $m_e$ is the electron mass, $c$ is the speed of light, $f_i$ is the oscillator strength of the $i^{th}$ transition, and $F_{\nu, i}$ is the stellar flux at the wavelength of transition $i$. The gravitational force is given by 

\begin{equation} \label{eq:gravity}
F_\mathrm{grav} = \frac{GM_{*}m_\mathrm{atom}}{d^2},
\end{equation}

\noindent where G is the gravitational constant, $M_{*}$ is the stellar mass, $m_\mathrm{atom}$ is the atomic mass, and $d$ is the distance of the atom from the star. The expression for $\beta$ is simply the ratio of Equations~\ref{eq:radiationpressure} and \ref{eq:gravity}: 

\begin{equation} \label{eq:beta}
\beta = \frac{1}{4\pi \epsilon_0}\frac{\pi e^2}{m_e c^2 } \frac{d^2}{GM_{*}m_\mathrm{atom}} \sum_{i=1} f_i F_{\nu,i}.
\end{equation}

\noindent Because the synthetic spectrum gives surface flux and therefore has a radius-squared dependency, the stellar surface gravity provides sufficient information for the $\beta$ calculation. We assume log $g_* = 4.2$ based on expectations for A0V stars\footnote{\url{http://www.pas.rochester.edu/~emamajek/EEM_dwarf_UBVIJHK_colors_Teff.txt}} \citep{Pecaut2013}, and note that this value is similar to the log $g_* = 4.3$ value derived from high-resolution optical spectra of \etaTel\ \citep{Rebollido:2018}. We adopt the smaller value to be consistent with the methodology used for other spectral types in Section~\ref{subsec:spectraltype_beta}. Our calculated $\beta$~values for \etaTel\ are listed in Table~\ref{table:lines}, in parentheses to the right of each species label. We find that all atomic species except \NI\ and \SII\ have $\beta > 0.5$; in contrast to the situation for the A5V star $\beta$~Pic, carbon and oxygen are not bound to the hotter A0V star \etaTel.

Note that in our presented $\beta$ values, we assume that only the $E_l = 0$ cm$^{-1}$ levels are populated, and our results are not sensitive to this assumption. In the limiting case where we assume all levels with $E_l <$ 500 cm$^{-1}$ are uniformly populated, the $\beta$ values of \CI, \CII, \OI, \SiII, \SI, and \FeII\ increase by factors of 1.7-3$\times$. Under either assumption, $\beta$ $>$ 0.5 for these species.  

Compared to the A0V $\beta$ calculations from \cite{Fernandez:2006} (see their Table 4), this work's $\beta$ values for \FeII, \SiII, \MgII, \SI, \SII, and \AlII\ are 1.3-10$\times$ larger, and our \CI\ and \CII\ values are larger by 17$\times$ and 190$\times$, respectively. For \CII, the braking agent in the $\beta$ Pic and 49 Cet debris disks, this difference is critically important as it is the only species where \cite{Fernandez:2006} finds $\beta < 0.5$ while this work finds $\beta \gg 0.5$. Differences in $\beta$ of 1.3-10$\times$ can be accounted for in the different synthetic spectra and atomic data used between the two works. The larger differences in the carbon $\beta$ values may be explained by where the dominant transitions lie in the stellar spectrum. For \CII, we find that the 1334.5 \AA\ transition dominates the radiation pressure by many orders of magnitude. The stellar flux densities between approximately 1200-1700 \AA\ are highly sensitive to the stellar effective temperature in the 9000-10,000 K range, and so small changes between the stellar properties used in this work and \cite{Fernandez:2006} might explain the drastically different carbon $\beta$ values. Although we selected the Castelli-Kurucz synthetic spectrum that most closely matched our STIS spectrum over all observed wavelengths, the low spectral resolution of the synthetic spectrum does not account for the presence of deep photospheric \CII\ absorption lines, even for a rapidly rotating star like \etaTel. In the vicinity of 1335 \AA, the $T_{eff}$ = 9750 K synthetic spectrum overestimates the true flux density as measured at high spectral resolution by STIS by a factor of three. We do not correct for this because $\beta$ would remain $\gg$0.5 and we wished to remain consistent in our use of synthetic spectra for $\beta$ calculations for other spectral types (Section~\ref{subsec:spectraltype_beta}).

To calculate $\beta$ for CO, we used molecular data from \citet{Morton1994} for UV transitions originating from the ground vibrational state ($v=0$) and HITRAN for transitons at $\lambda > 6000$~\AA\ \citep{Gordon2017}. Assuming thermal equilibrium, we calculated the fractional level populations and incorporated this into the calculation in Equation~\ref{eq:radiationpressure}. We found that $\beta$ is $\sim 11.6$ for temperatures below 20~K and decreases to 2.7 at $T = 500$~K, indicating any CO gas is not bound to the star.

\subsection{A Radiatively Driven Disk Wind} \label{subsec:diskwind}

\begin{figure*}
\includegraphics[width=\textwidth]{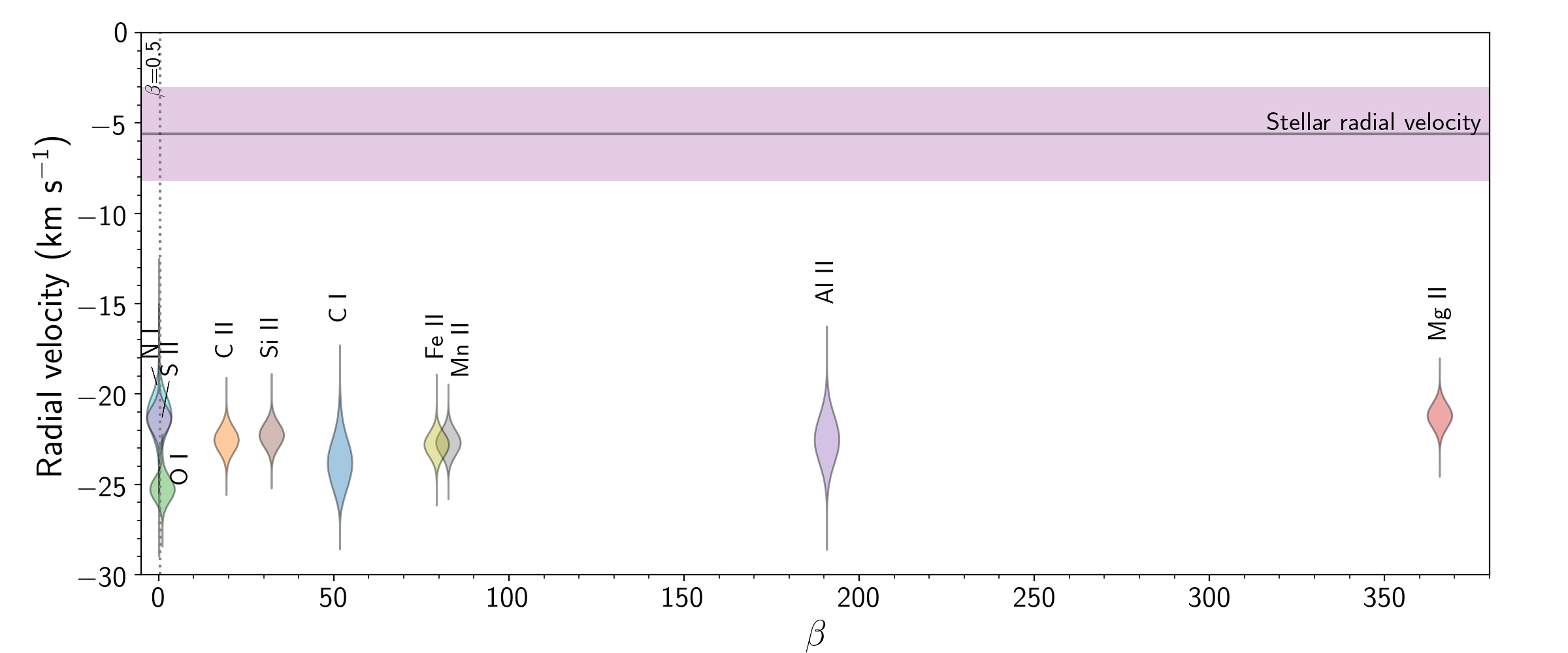}
\caption{The $\beta$~parameter (ratio of radiation pressure to gravitational force) compared to the fitted radial velocities of the CS absorbers. The violin plot illustrates the kernel probability density for each measurement, i.e., the width of the shaded area represents the proportion of the data located there. Uncertainties on the $\beta$~values are not shown, but are likely around a factor of 2-10$\times$. The vertical dotted line shows $\beta = 0.5$; species with $\beta > 0.5$ are not bound to the star. The stellar radial velocity from this work is shown with a black solid horizontal line and its uncertainty with the purple shaded region. For the CS and stellar radial velocities, the relative, rather than absolute, STIS wavelength calibration accuracy value (0.66 \kms) value has been propagated into the probability densities and uncertainties.} \label{fig:beta_comparison}
\end{figure*}

In Figure~\ref{fig:beta_comparison}, we compare the radial velocities of the CS absorbers (Table~\ref{table:lines}) to their corresponding $\beta$~parameter. We find that all of the species have a similar RV for their CS components ($-22.7\pm0.5$~\kms) despite a wide range of $\beta$~values from 0.05 for \SII\ to 365.7 for \MgII. All of the CS absorbers are blueshifted with respect to the star by $-16.9\pm2.6$~\kms, indicating that they are not bound to the star and are escaping in a disk wind. Note that the CS absorbers' average radial velocity in the stellar rest frame was computed with the STIS relative, rather than absolute, wavelength accuracy value propagated into the measured velocities from Table~\ref{table:lines} and Figure~\ref{fig:RV}. Even \NI\ and \SII, the only detected CS species with $\beta < 0.5$, are outflowing. This indicates that the \etaTel\ CS gas is acting as a single fluid with a mean abundance-weighted $\beta > 0.5$. This is similar to $\beta$~Pic \citep{Fernandez:2006}, except in the case of $\beta$~Pic, the mean $\beta$ of the fluid is less than 0.5 and the CS gas is bound to the star. 

The key to \CII\ acting as a braking agent in the case of $\beta$~Pic is that Coulomb forces increase the effective collisional cross-section for ions enough that the ionized gas can couple into a single fluid. Since \CII\ has $\beta < 0.5$ in the $\beta$~Pic CS environment, an overabundance of carbon can lower the mean abundance-weighted $\beta$ of the ionic fluid enough that it remains bound. Like $\beta$~Pic and 49~Cet, \etaTel\ is overabundant in carbon relative to Fe. However, unlike those systems, \etaTel\ has significantly more flux at the carbon transitions with the largest absorption oscillator strengths; this increases the \CI\ $\beta$ value to 51.7 and the \CII\ $\beta$ value to 19.3. Therefore, carbon is unable to reduce the mean abundance-weighted $\beta$ of the \etaTel\ CS fluid below 0.5; no amount of C in this gas will bind it to the star. 

Since \NI\ and \SII\ have $\beta < 0.5$, we consider whether these species \emph{could} act as a braking agent in the case of \etaTel, although they obviously do not. Ion-induced polarization of neutral particles increases their collisional cross section by orders of magnitude \citep{Beust1989,Fernandez:2006}, enabling gases like \NI\ to act as a braking agent. \NI\ is outflowing with the rest of the CS gas, indicating that it is coupled to the ionic gas through Coulomb forces or simply because the CS gas density is high enough. What overabundance of \NI\ or \SII\ would be required for these species to brake the whole \etaTel\ fluid remains a question for future dynamical modeling.

\section{Discussion} \label{section:discussion}

\subsection{Dynamics \& Composition in \etaTel, $\beta$~Pic, and 49~Cet}

The A0V spectral type of the \etaTel\ central star has a strong impact on the dynamics of the CS gas in this debris disk, producing an outflowing disk wind instead of the stable, bound gas disk seen around the A5V star $\beta$~Pic and the A1V star 49~Cet. Ironically, an outflowing disk wind is actually what was expected for $\beta$~Pic and the stability of the gas disk was puzzling for many years \citep[e.g.,][]{Lagrange:1998}. The extreme overabundance of carbon in the $\beta$~Pic and 49~Cet gas disks is necessary to bind them to the central stars.

This raises the question whether the overabundance of carbon in the $\beta$~Pic and 49~Cet CS gas reflects nature or nurture, meaning whether the gas is produced with some initial overabundance of carbon (preferential production) or whether the overabundance evolves from a gas with initial solar composition by radiative blow-out of heavier elements (preferential depletion). The detailed abundance pattern in the $\beta$~Pic gas tends to favor preferential production \citep{Xie:2013,Wilson2019}. In the case of \etaTel, the fact that all observed gas species are entrained in the radiatively driven disk wind, with no apparent dependence on $\beta$, qualitatively favors preferential production to explain the overabundance of volatiles and lithophiles relative to Fe. Alternatively, under-production of iron and other siderophiles might be causing an apparent overabundance of other elements.

\subsection{Impact of Spectral Type on Debris Disk Winds} \label{subsec:spectraltype_beta}

\begin{figure*}
\includegraphics[width=\textwidth]{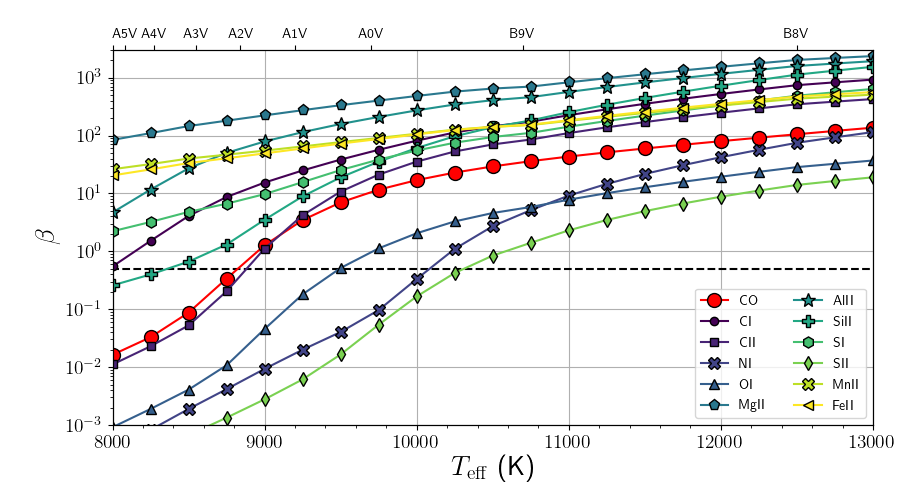}
\caption{The $\beta$ parameter as a function of stellar effective temperature and spectral type is shown for the molecular and atomic species discussed in this work. For CO, a Boltzmann distribution with excitation temperature $T_{ex}$ = 10 K was assumed. The horizontal dashed line shows $\beta = 0.5$.} \label{fig:beta_Teff}
\end{figure*}

To examine the effect of spectral type on the presence or absence of a disk wind more closely, we calculated the radiation pressure coefficients for a range of species as a function of the stellar effective temperature (Figure~\ref{fig:beta_Teff}), using the approach described in Section~\ref{sub:rad}. For this analysis, we only considered spectral types A5V and earlier, where radiative equilibrium synthetic spectra that do not include chromospheric emission generally agree with observed spectra at UV wavelengths. We estimate effective temperature, mass, and radius from \cite{Pecaut2013} \footnote{\url{https://www.pas.rochester.edu/~emamajek/EEM\_dwarf_UBVIJHK_colors_Teff.txt}}. The plot shows that at $T_\mathrm{eff}$ above about 10\,200~K (between A0V and B9V), all of the considered species have $\beta > 0.5$ and cannot brake a gaseous fluid, no matter how abundant. We should therefore not expect to see bound gas in debris disks around stars with earlier spectral types, but rather disk winds. This expectation is consistent with observation of a low-velocity wind from the debris disk around the B9V star $\sigma$~Herculis \citep{Chen:2003}.

This is not to say that debris disks around stars with spectral types later than about A0V/B9V 
must have bound stable gas. The presence or absence of a radiatively driven disk wind critically depends on the exact gas composition, as demonstrated in the case of $\beta$~Pic. Furthermore, for stars with spectral types later than about A5V, non-photospheric emission from stellar activity will come into play. The radiation pressure on a particular species depends on the exact stellar fluxes at the wavelengths of the absorbing transitions. One might therefore think that discrete stellar emission lines will not significantly impact the radiation pressure on a particular species. However, stellar emission lines involve many of the same transitions as the CS gas absorption lines. It is likely that a strong absorption line for a particular species will in fact overlap in wavelength with a stellar emission line, greatly increasing $\beta$. Also, stellar wind pressure may dominate over radiation pressure for late-type stars (e.g., \citealt{Sezestre2017}). Therefore, there may be a limited range of spectral types outside of which disk winds dominate, although estimating the range is beyond the scope of this work.

Figure~\ref{fig:beta_Teff} also indicates that \CII\ braking becomes ineffective at stellar effective temperatures above about 8800~K (A2V). This seems to indicate that \CII\ braking cannot explain the presence of a bound stable gas disk around the A1V star 49~Cet. However, difficulties in accurately modeling fluxes from real stars at the critical UV wavelengths and uncertainties in the atomic data itself lead to uncertainty in the exact threshold spectral type. A better understanding of the dynamics of the 49~Cet gas disk awaits abundance measurements for additional species in the gas (\SII\ appears critically important) and detailed dynamical modeling.

As is the case for \CII, stellar radiation pressure begins to dominate over gravity for CO above stellar effective temperatures of 8800~K (A2V). \cite{Roberge2000} found that CO is bound to the $\beta$ Pic (A5V) debris disk. This is consistent with our calculations that show $\beta \sim 0.02$ for A5V stars. Stable CO does not require \CII\ braking in the $\beta$ Pic disk, a process that is not as efficient as for the ions anyway \citep{Brandeker2011}. For \etaTel, any CO molecules should be blown out by the star, but the high stellar luminosity responsible for increasing $\beta$ also decreases CO's photoionization lifetime, meaning that detectable levels of CO cannot accumulate \citep{Kral2017}.

\section{Conclusion} \label{sec:conclusion}

The \textit{HST}/STIS UV spectra of \etaTel\ contribute to the growing inventory of atomic gas in debris disks and highlight the coupled effects of stellar type and composition on the gas dynamics. Here we summarize our key results.

\begin{itemize}

\item Absorption lines arising from \CI, \CII, \NI, \OI, \MgII, \AlII, \SiII, \SII, \MnII, and \FeII\ are detected in the spectra. The bulk of the gas is circumstellar rather than interstellar.

\item No CO absorption along the line of sight to the central star is detected in the UV spectra. A new upper limit on pure rotational CO emission in ALMA data is presented, but is relatively insensitive.

\item The CS gas appears to be overabundant in volatiles and siderophiles relative to iron. However, the C/O ratio is consistent with the solar ratio.

\item All the CS gas is blueshifted with respect to the central star and appears at a similar radial velocity regardless of the radiation pressure coefficient for each species. The CS gas dynamically behaves as a single fluid and is blowing out as a disk wind.

\item Compared to $\beta$~Pic (A5V), the hotter \etaTel\ (A0V) increases the radiation pressure coefficient for \CII\ and prevents this species from braking the fluid despite the overabundance of C relative to Fe.

\item Calculations of radiation pressure coefficients as a function of stellar effective temperature suggest that above about 10\,200~K (A0V to B9V) bound gas should not occur in debris disks, but rather disk winds. 

\end{itemize}

This work points to several avenues for exploration while forwarding our general understanding of gas in debris disks. Detailed dynamical modeling of the \etaTel\ gas should reveal whether preferential production explains the non-solar abundances. Similar modeling for 49~Cet is needed to identify the exact mechanism keeping the gas bound to the central star. A comprehensive investigation of the effect of spectral type on debris gas dynamics, with careful treatment of the stellar fluxes including chromospheric emission, could reveal which stars should show disk winds and which can have bound gas disks. Finally, sensitive UV spectra of more edge-on debris disks is needed to shed light on the detailed gas abundance patterns and what they might mean for the composition of the planetary material being destroyed to produce the gas.

\acknowledgments

A.Y. acknowledges support by an appointment to the NASA Postdoctoral Program at Goddard Space Flight Center, administered by USRA through a contract with NASA. M.A.M. acknowledges support from a National Science Foundation Astronomy and Astrophysics Postdoctoral Fellowship under Award No. AST-1701406. S.P. acknowledges support ANID/FONDECYT Regular grant 1191934. We thank Seth Redfield for helpful comments about the ISM analysis.
This research is based on observations made with the NASA/ESA \emph{Hubble Space Telescope} obtained from the Mikulski Archive for Space Telescopes (MAST) at the Space Telescope Science Institute, which is operated by the Association of Universities for Research in Astronomy, Inc., under NASA contract NAS 5–26555. These observations are associated with program HST-GO-14207 (PI: A. Roberge). The specific observations analyzed can be accessed via \dataset[10.17909/t9-8q9d-6c22]{https://doi.org/10.17909/t9-8q9d-6c22}. 
This paper also makes use of the following ALMA data: ADS/JAO.ALMA \#2013.1.01147.S (PI: S. P\'erez). ALMA is a partnership of ESO (representing its member states), NSF (USA), and NINS (Japan), together with NRC (Canada), MOST and ASIAA (Taiwan), and KASI (Republic of Korea), in cooperation with the Republic of Chile. The Joint ALMA Observatory is operated by ESO, AUI/NRAO and NAOJ. The National Radio Astronomy Observatory is a facility of the National Science Foundation operated under cooperative agreement by Associated Universities, Inc. 
This research has made use of NASA's Astrophysics Data System Bibliographic Services and the SIMBAD database, operated at CDS, Strasbourg, France.

\facilities{HST, ALMA}
\software{astropy \citep{Robitaille2013}, IPython \citep{Perez2007}, Matplotlib \citep{Hunter2007}, NumPy and SciPy \citep{VanderWalt2011}, lyapy \citep{Youngblood2016}, emcee \citep{Foreman-Mackey:2013}, CASA \citep{McMullin:2007}, PyAstronomy\footnote{\url{https://github.com/sczesla/PyAstronomy}} \citep{pya}, molecular-hydrogen\footnote{\url{https://github.com/keflavich/molecular\_hydrogen}}.}

\bibliography{main.bbl}{}
\bibliographystyle{aasjournal}

\end{document}